\documentclass[journal]{IEEEtran}
\usepackage{setspace}
\usepackage[utf8]{inputenc}
\usepackage{cite,graphicx,amsmath,amssymb}
\usepackage{hyperref}
\usepackage{subfigure}
\usepackage{fancyhdr}
\usepackage{mdwmath}
\usepackage{tabularx}
\usepackage{mdwtab}
\usepackage{balance}
\usepackage{xcolor}
\usepackage{bm}
\usepackage{amsthm}
\usepackage{threeparttable}
\usepackage{multirow}
\usepackage{amsmath}
\usepackage{flafter}
\usepackage{makecell}
\usepackage{diagbox}
\usepackage{tcolorbox}
\usepackage{physics}
\usepackage{bbding}
\usepackage{pifont}
\usepackage{algorithm}
\usepackage{algorithmic}
\usepackage{booktabs}
\usepackage{mathtools}
\usepackage{enumitem}
\usepackage{subcaption}
\usepackage{array}
\usepackage{ragged2e}
\newcolumntype{L}[1]{>{\RaggedRight\arraybackslash}p{#1}}

\hypersetup{colorlinks=true,
linkcolor=blue,
citecolor=blue,      
urlcolor=blue,
}

\theoremstyle{definition}

\newtheorem{theorem}{Theorem}

\newtheorem{lemma}{Lemma}

\newtheorem{corollary}{Corollary}

\newcolumntype{L}{>{\raggedright\arraybackslash}X}
\makeatletter

\begin{document}
\bstctlcite{IEEEexample:BSTcontrol}
\title{A Survey of Pinching-Antenna Systems (PASS)}
\author{Yuanwei Liu, \IEEEmembership{Fellow, IEEE}, Hao Jiang, Xu Gan, Xiaoxia Xu, Jia Guo,  Zhaolin Wang, Chongjun Ouyang, Xidong Mu, Zhiguo Ding, \IEEEmembership{Fellow, IEEE}, Arumugam Nallanathan, \IEEEmembership{Fellow, IEEE}, Octavia A. Dobre, \IEEEmembership{Fellow, IEEE}, \\ George K. Karagiannidis, \IEEEmembership{Fellow, IEEE}, and Robert Schober, \IEEEmembership{Fellow, IEEE}
% \thanks{(Corresponding author: Yuanwei Liu; Xidong Mu)}
\thanks{Y. Liu, X. Gan, and Z. Wang are with the Department of Electrical and Electronic Engineering, The University of Hong Kong, Hong Kong (e-mail: \{yuanwei, eee.ganxu, zhaolin.wang\}@hku.hk).}
\thanks{H. Jiang, J. Guo, X. Xu, C. Ouyang, and A. Nallanathan are with the School of Electronic Engineering and Computer Science, Queen Mary University of London, London, E1 4NS, U.K. (e-mail: \{hao.jiang, jia.guo, x.xiaoxia, c.ouyang, a.nallanathan\}@qmul.ac.uk).}
\thanks{X. Mu is with the Centre for Wireless Innovation (CWI), Queen's University Belfast, Belfast, BT3 9DT, U.K. (e-mail: x.mu@qub.ac.uk).}
\thanks{O. A. Dobre is with the Faculty of Engineering and Applied Science, Memorial University, St. John’s, NL A1C 5S7, Canada. (e-mail: odobre@mun.ca)}
\thanks{Z. Ding is with the School of Electrical and Electronic Engineering (EEE), Nanyang Technological University, Singapore 639798 (e-mail: zhiguo.ding@ntu.edu.sg).}
\thanks{G. K. Karagiannidis is with the Wireless Communications and
Information Processing (WCIP) Group, Electrical \& Computer Engineering
Dept., Aristotle University of Thessaloniki, 54124, Thessaloniki, Greece (e-mail: geokarag@auth.gr).}
\thanks{R. Schober is with the Institute for Digital
Communications, Friedrich-Alexander-University Erlangen-Nurnberg (FAU), 91054 Erlangen, Germany (e-mail:
robert.schober@fau.de).}
}
\maketitle
% Authors
\begin{abstract}
The pinching-antenna system (PASS), recently proposed as a flexible-antenna technology, has been regarded as a promising solution for several challenges in next-generation wireless networks. It provides large-scale antenna reconfiguration, establishes stable line-of-sight links, mitigates signal blockage, and exploits near-field advantages through its distinctive architecture. This article aims to present a comprehensive overview of the state of the art in PASS. The fundamental principles of PASS are first discussed, including its hardware architecture, circuit and physical models, and signal models. Several emerging PASS designs, such as segmented waveguide-enabled PASS (SWAN), center-fed PASS (C-PASS), and multi-mode PASS (M-PASS), are subsequently introduced, and their design features are discussed. In addition, the properties and promising applications of PASS for wireless sensing are reviewed. On this basis, recent progress in the performance analysis of PASS for both communications and sensing is surveyed, and the performance gains achieved by PASS are highlighted. Existing research contributions in optimization and machine learning are also summarized, with the practical challenges of beamforming and resource allocation being identified in relation to the unique transmission structure and propagation characteristics of PASS. Finally, several variants of PASS are presented, and key implementation challenges that remain open for future study are discussed.
\end{abstract}

\section{Introduction}
With the fifth-generation (5G) communication technology commercialized globally, a persistent question arises: \emph{What will the next-generation communication networks be like?} 
Building upon the solid foundation of 5G, next-generation telecommunication networks are anticipated to not only advance beyond current 5G capabilities but also to explore untapped areas, heralding a world with high-speed coverage, intelligent perception, and massive connectivity \cite{jiang2021road, giordani2020toward, kaushik2024toward, tataria20216g}.

To realize these ambitious goals, multiple-input multiple-output (MIMO) technology is expected to remain a key enabler of revolutionary changes \cite{busari2018millimeter, yang2015fifty}. 
In fact, MIMO technology has had a profound impact on the evolution of wireless communications over the past decades, known for its ability to offer enhanced array gains, multiplexing gains, and diversity gains \cite{new2024fluid}. 
In recent years, MIMO has evolved into multiple advanced forms, such as holographic MIMO (H-MIMO) and extremely large-scale antenna arrays (ELAAs), thereby offering multiple avenues to meet the ever-increasing demands of research, standardization, and commercial applications. 

From a conventional MIMO perspective, wireless channels are treated as fixed parameters determined by site-specific propagation conditions. 
Therefore, to ensure reliable transmission, higher hardware costs are unavoidable to overcome undesirable propagation conditions. To address this issue, NTT DOCOMO introduced the pinching-antenna system (PASS) in 2022 \cite{suzuki2022pinching, pinching_antenna1}. 
Unlike conventional fixed-position MIMO, PASS consists of multiple low-attenuation waveguides that extend tens of meters. Each waveguide is attached to one or more radiation particles, known as pinching antennas (PAs). 
Leveraging the flexibility of positioning PAs at the meter scale allows wireless signals to radiate freely from the waveguides. This endows PASS with the ability to proactively reconfigure wireless channels \cite{liu2025pinch_mag}.
With this flexibility, PASS can be viewed as the outer layer of a tri-hybrid beamforming paradigm for next-generation beamforming configuration \cite{castlanos2025embracing}, featuring: \emph{i)~Line-of-Sight (LoS) Creation}, {\emph{ii)~Near-Field Benefits}}, and {\emph{iii)~Low-Cost and Scalable Implementation}} \cite{liu2025pinch_mag, liu2025pinching}.

\subsection{Flexible Antennas}
Known as one of the most important driving forces in the development of telecommunications, MIMO has been extensively investigated, ranging from theoretical analysis \cite{sohrabi2016hybrid} to real-world implementations \cite{emil2025enabling}.
In general, there are three main technical trends for MIMO: \emph{i) Larger Aperture Size}, \emph{ii)~Denser Antenna Placement}, and \emph{iii) More Flexible Antenna Architectures}.

The first trend stems from the fact that increasing the number of antenna elements enhances beamforming gain \cite{liu2023near_tut}.
In particular, since the high-frequency bands, such as the millimeter-wave or terahertz bands, are exploited for their abundant bandwidth, large beam gains play a vital role in overcoming the severe atmosphere-induced attenuation \cite{jiang2023active, jiang2024sense}.
From a multiplexing perspective, a large number of antennas creates high-dimensional channel vectors, ensuring the orthogonality among users and suppressing inter-user interference.
These benefits and the higher carrier frequency motivate the research community to integrate more antennas at base stations, which, in turn, leads to more pronounced near-field phenomena \cite{cui2023near}. 
As such, antenna arrays can function like ``lenses'' to focus beams to a particular spot rather than simply steering them in a particular direction.
To realize the benefits mentioned above, antenna arrays are undergoing a transformation, from massive MIMO to gigantic MIMO~\cite{emil2025enabling}.

Unlike the first trend of scaling up arrays by deploying more elements while adhering to the half-wavelength rule, the second trend of densifying arrays does not necessarily follow this rule for antenna elements. In the context of holographic MIMO, more antenna elements are integrated into a given aperture size. In fact, a perfectly continuous aperture, known as a continuous aperture array (CAPA), can be formed \cite{liu2025capa, wang2025beamforming}. 
Therefore, rather than designing phase shifts and amplitudes for a large number of discrete antenna elements, holographic MIMO allows for analysis and optimization of a continuous source-current function. 
By precisely manipulating the source current, radiation patterns or electromagnetic (EM) properties can be precisely shaped, thus leading to enhanced performance, as confirmed by references \cite{zhang2023pattern} and \cite{pizzo2022fouier}. 
Furthermore, holographic MIMO offers energy- and cost-efficient solutions because the number of power-intensive RF chains is determined by the number of spatially multiplexed data streams rather than the number of antennas.

Although these trends have demonstrated their great potential, there are non-negligible challenges along the path to real-world implementations. 
For larger-aperture arrays, the primary obstacle is the increased signal-processing overhead. 
Specifically, theoretical performance gains are realized only when beamforming vectors are effectively optimized. 
However, the time and computational resources required to manipulate these variables effectively can be formidable, especially in the ELAA case, since the number of optimization variables is proportional to the number of antenna elements, which can number in the hundreds or even thousands \cite{ye2024extremely}. 
This could result in intolerable communication overhead and hinder real-time communication. 
For densified arrays, although optimization overhead can be reduced by optimizing a continuous function, mutual coupling induced by close antenna placement acts as a limiting factor and requires significant attention \cite{pizzo2025mutual}. 
Additionally, their performance is passively determined by the propagation conditions. 
In other words, these MIMO variations lack the ability to reconfigure the wireless channel.

To compensate for this limitation, flexible antennas, i.e., the third technical trend, are devised to introduce wireless channel reconfigurability, i.e., the ability to strategically manipulate the wireless channel to meet specific requirements \cite{zhu2026movable, ding2025flexible}.
From a historical perspective, flexible antennas have a long history.
In the third-generation (3G) era, antenna flexibility was achieved through antenna selection, which exploits spatial diversity by selecting antennas with desirable propagation conditions and achieves a good balance between performance and hardware cost \cite{molish2004mimo, sanayei2004antenna}.
Subsequently, reconfigurable intelligent surfaces (RISs) have emerged to flexibly reconfigure wireless channels, offering the benefits of virtual LoS links between transceivers, thereby artificially tuning wireless channels \cite{liu2021reconfigurable, wu2019intelligent}.
Recently, movable and fluid antenna systems have been introduced into the realm of flexible antenna technologies, which aim at liberating the antenna elements from conventional fixed-position layouts \cite{hong2025contemporaray, new2025fluid, zhu2024movable_mag, zhu2025tutorial}.
These new forms of antennas allow changes in antenna element position or shape, thereby modifying wireless channel characteristics.
However, whether fluid or movable antennas are considered, only antenna repositioning on a scale of a few wavelengths is supported, thereby making them suitable for mitigating small-scale fading. 
In fact, wavelength-scale motions are sufficient in rich-scattering cases, since the gains of the links fluctuate significantly over the entire aperture size.
However, in high-frequency bands dominated by LoS links, flat channel conditions across the entire array render spatial diversity unavailable.

To mitigate this issue, several attempts have been made to explore reconfiguration to combat large-scale fading.
For instance, surface wave communication (SWC) enables signal propagation to occur at the surface between the air and the metal ground rather than free space, so as to reduce the large-scale path loss \cite{chu2025on}.
Additionally, an extremely large movable antenna was proposed in \cite{fu2025extremely} to enable antenna movements over areas of hundreds of square meters, leading to effective reconfiguration of large-scale links.

\subsection{Pinching-Antenna System (PASS)}
Although current flexible antenna technologies have demonstrated significant potential for channel reconfigurability, they face two primary limitations: Limited flexibility and lack of scalability.
Specifically, most flexible antennas permit repositioning on a wavelength scale, making them non-viable for manipulating large-scale fading.
For the minority that permits large-scale antenna flexibility, deployment costs remain high. 
Moreover, once deployed, adding or removing antennas/ports often incurs substantial additional costs and may be difficult to realize in many installations.

As a remedy, PASS can address these issues while maintaining the flexibility to reconfigure both large- and small-scale channel characteristics.
By exploiting signal propagation within waveguides, PASS can reduce signal power dissipation due to the low attenuation in the waveguide.
At desirable positions, PAs are installed to radiate the guided signal into free space, and short-range free-space propagation is used as the final step to deliver the signal to the users.
This radiation step is supported by coupling theory \cite{wang2025multiport, wang2025modeling}.
Through judicious design, e.g., antenna repositioning, in-waveguide and free-space propagation can be integrated to facilitate transmission.
More importantly, repositioning PAs can proactively adjust channel state information (CSI) and even create strong LoS conditions. 
In addition to the above, the low cost of waveguides and the ease of attaching and detaching PAs enable low-cost implementation and higher scalability than other forms of flexible antennas.

From a historical perspective, the concept of PAs originates from the development of leaky coaxial cables (LCXs), in which fixed-position slots are drilled into coaxial cables to radiate guided signals \cite{xu2025generalized, jun2001theory, ziying2024theoretical}.
Due to the flexibility of waveguides, LCXs have been widely used in narrow, sealed spaces, such as tunnels and mines, serving as a solid hardware foundation for PASS \cite{liu2025pinch_mag}.
In contrast to the inflexibility of slots fixed on LCXs, PASS comprises low-attenuation waveguides extending over tens of meters, with single or multiple PAs attached to each waveguide that can be flexibly repositioned.
Inspired by a pioneering paper \cite{ding2025flexible}, a plethora of works have been dedicated to revealing the potential of PASS in wireless communications, encompassing optimization \cite{wang2025modeling, bereyhi2025mimo}, performance analysis \cite{ouyang2025array, ding2025analytical, gu2026optimal}, multiple access \cite{ding2025los, zhao2026waveguide, shan2026access}, and a wide spectrum of applications; see also \cite{liu2023near_tut}.
Furthermore, several new variants of PASS have been introduced, such as segmented structures \cite{ouyang2025uplink}, center-fed layouts \cite{gan2025c}, and wireless feeding approaches \cite{wijewardhana2025wireless}, among others.
To further exploit spatial flexibility, a two-dimensional planar PASS was introduced in \cite{zhong2025two}, in which the PAs can be repositioned over a two-dimensional plane. This novel architecture provides additional spatial DoFs, thereby enhancing the performance of PASS and offering greater gains than conventional one-dimensional PASS configurations.
In parallel, transmit PASS (T-PASS) was proposed by the authors in \cite{gan2026T_pass}, in which the guided energy is reused by a wired user. 
As summarized in Table~\ref{tab:comparison_pass_mimo}, there are multiple aspects in which PASS outperforms the conventional antennas.
In particular, in terms of flexibility, PASS allows both large- and small-scale antenna repositioning \cite{liu2025pinch_mag, liu2025pinching}, thereby enabling reconfiguration of wireless propagation conditions.
However, conventional antennas are fixed and cannot be reconfigured. 
In terms of multiple access domains, PASS can not only enable multiplexing through conventional domains, such as time, frequency, and spatial, but also allow new multiple-access dimensions, such as waveguide \cite{shan2026multigroup, zhao2026waveguide} and wireless environment \cite{ding2026environment, xu2026pinching}.
Furthermore, from a communication perspective, PASS employs pinching beamforming on top of a conventional two-layer hybrid beamforming architecture, namely tri-hybrid beamforming, to further enhance throughput \cite{jiang2026exploiting,zhao2025tri,guo2026graph}. 
Finally, in terms of sensing, PASS can create a large aperture size and long-extending waveguides using a few PAs, thereby enhancing sensing resolution and full-dimensional sensing with near-field effects \cite{jiang2026pinching, liu2025near}.
On the contrary, conventional antennas can only use far-field one-dimensional sensing due to limited aperture sizes, or near-field sensing enabled by hundreds of antenna elements. 
% To further highlight the distinctive characteristics of PASS, Table~\ref{tab:comparison_pass_mimo} compares PASS with conventional antenna systems.
\begin{table*}[t] 
\centering
\caption{Comparison between conventional antenna systems and PASS.}
\label{tab:comparison_pass_mimo}
\setlength{\tabcolsep}{3pt}
\renewcommand{\arraystretch}{1.25}
\scriptsize
\resizebox{0.82\textwidth}{!}{%
\begin{tabular}{|c|c|c|}
\hline
\makecell[c]{\textbf{Aspect}} &
\makecell[c]{\textbf{Conventional Antenna}} &
\makecell[c]{\textbf{PASS}} \\
\hline
\makecell[c]{\textbf{Hardware}} &
\makecell[c]{Fixed antenna arrays connected to RF chains.} &
\makecell[c]{Dielectric waveguides \\ with reconfigurable PAs.} \\
\hline

\makecell[c]{\textbf{Flexibility}} &
\makecell[c]{Fixed antenna positions.} &
\makecell[c]{Large-scale PA-position reconfiguration.} \\
\hline

\makecell[c]{\textbf{Multiple Access}} &
\makecell[c]{Time, frequency, power,\\ and spatial domains.} &
\makecell[c]{Time, frequency, power, spatial,\\ PA-position, and waveguide domains.} \\
\hline

\makecell[c]{\textbf{Beamforming}\\\textbf{Architecture}} &
\makecell[c]{Digital and analog \\ beamforming.} &
\makecell[c]{Digital, analog, and \\ pinching beamforming.} \\
\hline

\makecell[c]{\textbf{Sensing}\\\textbf{Capability}} &
\makecell[c]{Far-field and near-field sensing \\ with a fixed antenna geometry.} &
\makecell[c]{Near-field sensing with a \\ reconfigurable PA geometry.} \\
\hline
\end{tabular}%
}
\end{table*}

It is worth clarifying the relationship between FAS and PASS, since both belong to the broader family of flexible antennas.
In particular, FAS was originally proposed to leverage antenna position flexibility within a compact aperture to select antenna ports with better channel conditions.
Consequently, the diversity, outage, and multiplexing performance of FAS-based communication systems can be enhanced~\cite{wong2021fluid,new2024fluid,wong2022fluid,wong2023opportunistic}.
In addition, some practical issues, such as channel estimation (CE) across multiple candidate ports, have also been investigated, revealing the trade-off between CSI acquisition overhead and port-selection performance \cite{skouroumounis2023fluid}.
Recent tutorial and survey papers have further summarized the potential of FAS for future communication networks~\cite{new2025fluid, hong2025contemporaray}.
However, the effectiveness of FAS is primarily achieved through wavelength-scale reconfiguration, as its aperture size is typically small.
Therefore, FAS is particularly effective in rich-scattering environments, where small position changes can lead to noticeable channel fluctuations.
In contrast, PASS exploits low-loss waveguide propagation and repositions PAs to create better channel conditions at meter-scale locations.
Consequently, PASS can reconfigure not only small-scale fading but also large-scale propagation characteristics, making it especially attractive for scenarios dominated by LoS links.
However, these advantages also come with distinct challenges.
While FAS requires compact reconfigurable hardware, fast port selection, and efficient CE, PASS necessitates careful waveguide deployment, PA coupling and radiation modeling, in-waveguide loss control, and effective PA repositioning methods.
In summary, FAS and PASS are complementary rather than competing technologies: FAS is better suited to rich scattering environments, whereas PASS is better suited to reconfiguring large-scale channels and LoS-link creation.

\subsection{Motivation and Contribution}
    Given the rapid advancement of PASS, a comprehensive survey summarizing current progress and outlining future research directions is warranted.
    Compared with magazine papers in \cite{yang2025pinch, liu2025pinch_mag, zeng2025resource}, this work presents a comprehensive review of the existing work on PASS.
    In parallel, the authors of \cite{yu2026intelligent} conducted a survey of multiple types of intelligent flexible-position antenna systems and provided only a concise discussion of PASS, while \cite{duan2025physical} examines PASS only from a near-field communication perspective.     
    Compared with tutorial papers in \cite{liu2025pinching, xu2025generalized, illi2026comprehensive}, this survey first offers an integrated perspective on practical PA implementation, hardware modeling, and newly emerging PASS architectures. 
    In particular, we present the small-dielectric-scatterer model, the directional-coupler-based waveguide model, and the multiport network model for PAs. 
    Compared with existing simplified PA modeling methods \cite{liu2025pinch_mag, liu2025pinching}, these models capture distinct physical mechanisms underlying signal coupling, radiation, and mutual interactions among PAs, thereby providing a more realistic foundation for the analysis and design of practical PASS.
    Moreover, this survey highlights several emerging PASS architectures, including segmented waveguide-enabled PASS, center-fed PASS, and multi-mode PASS, which are novel variants of conventional PASS \cite{pinching_antenna1}. 
    These architectures extend the conventional PASS, thereby enhancing its flexibility and enabling new designs and optimization techniques.
    
    A detailed comparison between this work and the existing papers is shown in Tab. \ref{tab:r2c1_comparison_existing_works}.
    \begin{table*}[h!]
        \caption{Comparison of this work with existing magazine/survey/tutorial papers on PASS. Here, $\bigstar$, $\blacksquare$, and $\blacktriangle$ refer to ``high-level overview'', ``detailed tutorial'', and ``mentioned but not considered in detail'', respectively.}
        \label{tab:r2c1_comparison_existing_works}
        \centering
        \footnotesize
        \setlength{\tabcolsep}{3pt}
        \renewcommand{\arraystretch}{1.1}
        \resizebox{\textwidth}{!}{%
        \begin{tabular}{|c|c|c|c|c|c|c|c|}
            \hline
            \textbf{Ref.} & \textbf{Type} & \textbf{Principles} & \textbf{Hardware}
            & \textbf{Communication Func.} & \textbf{Sensing Func.}
            & \textbf{Future Directions} & \textbf{Advanced Forms} \\
            \hline
            \cite{pinching_antenna1} & Demo. & $\bigstar$ & & & & & \\
            \hline
            \cite{suzuki2022pinching} & Mag. & $\bigstar$ & $\bigstar$ & $\blacktriangle$ & & & \\
            \hline
            \cite{liu2025pinch_mag} & Mag. & $\bigstar$ & $\bigstar$ & $\bigstar$ & $\blacktriangle$ & $\blacktriangle$ & \\
            \hline
            \cite{yang2025pinch} & Mag. & $\bigstar$ & $\blacktriangle$ & $\blacktriangle$ & & $\blacktriangle$ & \\
            \hline
            \cite{xu2025generalized} & Tut. & $\bigstar$ & $\blacktriangle$ & $\blacksquare$ & $\blacksquare$ & $\bigstar$ & $\blacktriangle$ \\
            \hline
            \cite{liu2025pinching} & Tut. & $\bigstar$ & $\blacksquare$ & $\blacksquare$ & $\blacktriangle$ & $\bigstar$ & \\
            \hline
            \cite{yu2026intelligent} & Surv. & $\bigstar$ & $\blacktriangle$ & & & & \\
            \hline
            \cite{duan2025physical} & Surv. & $\bigstar$ & $\blacktriangle$ & $\blacktriangle$ & & $\blacktriangle$ & \\
            \hline
            \cite{zeng2025resource} & Preprint & $\blacktriangle$ & $\blacktriangle$ & & & $\blacktriangle$ & \\
            \hline
            \cite{illi2026comprehensive} & Preprint & $\blacktriangle$ & $\blacktriangle$ & $\blacktriangle$ & $\blacktriangle$ & $\blacktriangle$ & $\blacktriangle$ \\
            \hline
            \textbf{This work} & Surv. & $\bigstar$ & $\bigstar$ & $\bigstar$ & $\bigstar$ & $\bigstar$ & $\bigstar$ \\
            \hline
        \end{tabular}%
        }
    \end{table*}
    
    {Given the background outlined above, the contributions of this paper are summarized as follows:
    \begin{itemize}
    \item We present the fundamental theories and modeling techniques for PASS. 
    In particular, the implementation principles of PASS are introduced through small-dielectric-scatterer-based PAs and the directional-coupler waveguide model. 
    Then, multiport theory is discussed, which provides a mathematically tractable framework for characterizing the electromagnetic behavior of guided signals. 
    Finally, based on these theoretical foundations, the PASS signal is elaborated.
        
    \item We further propose several advanced PASS architectures. 
    Specifically, segmented-waveguide-enabled PASS (SWAN) employs multiple shorter waveguide segments with dedicated feed points, which simplifies uplink channel modeling and mitigates in-waveguide propagation loss. 
    In parallel, center-fed PASS (C-PASS) incorporates a T-junction power splitter to enable more flexible signal routing within the waveguide. 
    Moreover, multi-mode multi-feed PASS (M-PASS) enables the simultaneous propagation of multiple modes, thereby enhancing the multiplexing capability of PASS.
        
    \item We further cover PASS-assisted sensing.
    In particular, a systematic review of PASS-assisted sensing is presented, covering sensing, localization, and user tracking.
    Next, the interplay between sensing and communication is unveiled from the perspectives of sensing-aided communications and communication for sensing, respectively.
        
    \item We review the performance analysis of PASS. From a communication perspective, existing studies are categorized according to achievable-rate and outage-probability analyses. From a sensing perspective, existing studies on sensing performance analysis are classified according to the sensing metrics employed.

    \item Finally, we review advanced mathematical tools for addressing the intrinsic challenges of PASS optimization. 
    From a conventional optimization perspective, existing methods are categorized into three groups: Structure-based optimization, population-based optimization, and game-theoretic optimization. 
    Furthermore, a machine-learning-based approach is presented to highlight the potential of integrating AI techniques into PASS optimization.
\end{itemize}}
\begin{table*}[!t]
\caption{List of Key Acronyms}
\label{tab:LISTOFACRONYMS}
\centering
\resizebox{0.82\textwidth}{!}{
\begin{tabular}{|l||l|l||l|}
\hline
PASS & Pinching-Antenna System & PA & Pinching Antenna \\ \hline
1D & One-Dimensional & 2D & Two-Dimensional \\ \hline
LoS & Line-of-Sight & NLoS & Non-Line-of-Sight \\ \hline
MIMO & Multiple-Input Multiple-Output & RF & Radio Frequency \\ \hline
EM & Electromagnetic & CSI & Channel State Information \\ \hline
SNR & Signal-to-Noise Ratio & SE & Spectral Efficiency \\ \hline
DoF & Degree of Freedom & BS & Base Station \\ \hline
SWAN & Segmented Waveguide-Enabled Pinching-Antenna System & C-PASS & Center-Fed Pinching-Antenna System \\ \hline
M-PASS & Multi-Mode Pinching-Antenna System & IAR & Inter-Antenna Radiation \\ \hline
ISAC & Integrated Sensing and Communications & CRLB & Cram\'er--Rao Lower Bound \\ \hline
BCRLB & Bayesian Cram\'er--Rao Lower Bound & OP & Outage Probability \\ \hline
ML & Machine Learning & GNN & Graph Neural Network \\ \hline
DNN & Deep Neural Network & NOMA & Non-Orthogonal Multiple Access \\ \hline
DL & Downlink & UL & Uplink \\ \hline
OFDMA & Orthogonal Frequency-Division Multiple Access & FPA & Fixed-Position Array \\ \hline
LCX & Leaky Coaxial Cable & DCW & Directional Coupler Waveguide \\ \hline
SDS & Small Dielectric Scatterer & TM & Transverse Magnetic \\ \hline
TE & Transverse Electric & & \\ \hline
\end{tabular}}
\end{table*}

\begin{figure}[t!]
    \centering
    \includegraphics[width=0.96\linewidth]{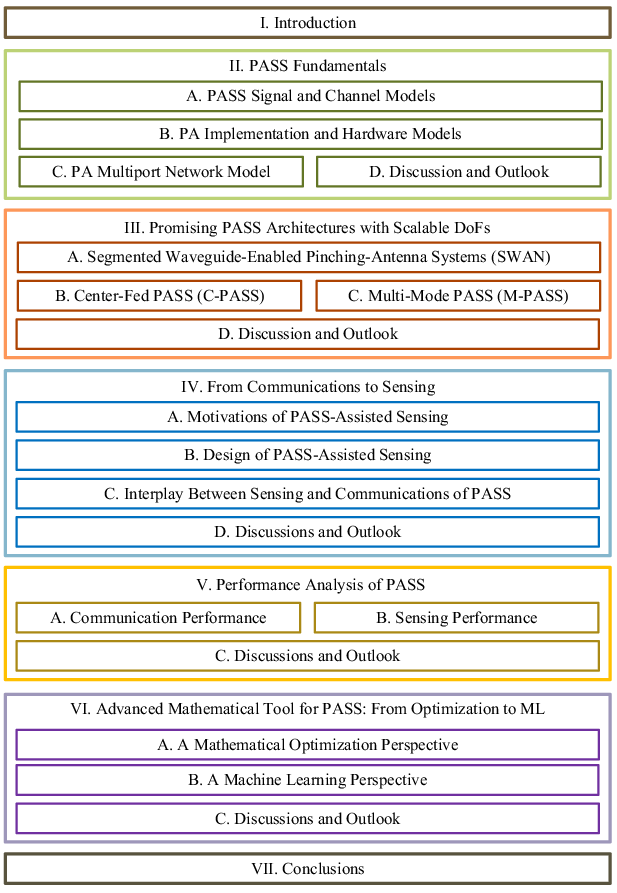}
    \caption{Condensed overview of this survey on PASS.}
    \label{fig:overview}
\end{figure}
\subsection{Notations and Organization}
The remainder of this paper is structured as follows:
Section \ref{sect:fundamentals} presents the physical models for PASS, followed by Section \ref{sect:promising}, which introduces several newly-emerged variations of PASS.
In the sequel, Section \ref{sect:communication_to_sensing} discusses the communication and sensing functionalities of PASS.
Section \ref{sect:preformance_analysis} offers a comprehensive survey on performance analysis of PASS, while Section \ref{sect:optimization} details the mathematical tools for PASS optimization.
Finally, the paper is concluded by Section \ref{sect:conclusions}.
Fig. \ref{fig:overview} illustrates the organizational structure
of the paper, and Table \ref{tab:LISTOFACRONYMS} presents key abbreviations used in this work.

\emph{Notations:} Throughout this paper, scalars, vectors, and matrices are denoted by lowercase, bold lowercase, and bold uppercase letters, respectively. The superscripts $(\cdot)^T$ and $(\cdot)^H$ denote transpose and Hermitian transpose, respectively. $\|\cdot\|$ denotes the Euclidean norm, $|\cdot|$ denotes the absolute value or cardinality depending on the context, and $\mathbb{E}\{\cdot\}$ denotes statistical expectation. $\mathbb{C}$ and $\mathbb{R}$ denote the sets of complex and real numbers, respectively.

\section{PASS Fundamentals}\label{sect:fundamentals}
In this section, we introduce the fundamentals of PASS for wireless communications. The basic signal and channel models of PASS are first introduced, followed by its implementation and hardware models, as well as the abstract multiport network models. 

\begin{figure}[t]
	\centering
	\includegraphics[width=0.8\linewidth]{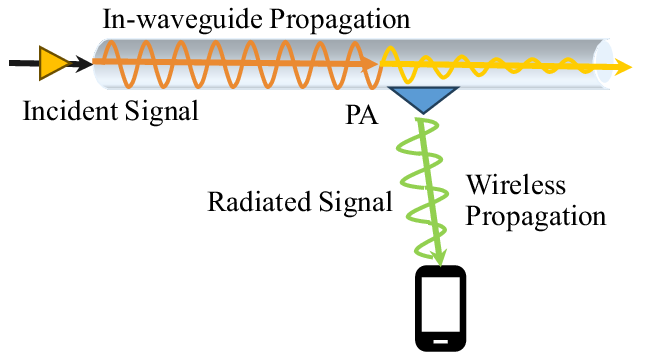}
	\caption{Illustration of signal propagation in PASS.}
	\label{fig:pa_radiate}
\end{figure}

\subsection{PASS Signal and Channel Models}
For the purpose of exposition, we first consider a simple downlink, single-waveguide, single-PA, single-user setup, as illustrated in Fig.~\ref{fig:pa_radiate}. The system is narrowband at a given carrier frequency. Therefore, the end-to-end channel between transceivers can be represented by a single channel tap.

The signal propagation in PASS naturally decomposes into two components: \emph{In-waveguide propagation} and \emph{free-space propagation}. These two stages are linked by the PA. Let $s_{\mathrm{in}}(t)$ denote the baseband equivalent of the signal injected into the waveguide at the feed point. The in-waveguide propagation describes how this signal evolves as it travels a distance $L$ along the waveguide before reaching the PA. Due to the finite conductor resistance and dielectric loss, the signal experiences attenuation and phase rotation. The relationship between the signal at the PA location, $s_{\mathrm{pa}}$, and the incident signal $s_{\mathrm{in}}$ can be characterized using the standard energy attenuation model \cite[Eq.~(8-47)]{cheng1989field} as $s_{\mathrm{pa}}(t) = \exp\left(-\gamma_g L\right) s_{\mathrm{in}}(t)$. Here, $\gamma_g$ is the complex propagation constant inside the waveguide, defined as $\gamma_g = \alpha_g + j \beta_g$, where $\alpha_g > 0$ and $\beta_g > 0$ are the amplitude attenuation and phase constants, respectively. Their values are determined by the waveguide geometry, material parameters, and the operating wavelength, and can be obtained from \cite[Eqs.~(8-48) \& (8-49)]{cheng1989field}. In particular, by defining $\lambda_g$ as the effective wavelength in the waveguide medium, $\beta_g$ can be expressed as $\beta_g = 2\pi/\lambda_g$~\cite{cheng1989field}, which is consistent with the in-waveguide signal model adopted in~\cite{ding2025flexible}. For an ideal lossless waveguide, we have $\alpha_g = 0$, and hence $\lvert s_{\mathrm{pa}}(t) \rvert = \lvert s_{\mathrm{in}}(t) \rvert$.

We now turn to the free-space propagation from the PA to the user. For exposition, we focus on a single line-of-sight (LoS) path. Let $R$ denote the LoS distance between the PA and the user. For this setting, the end-to-end baseband channel coefficient for a lossless waveguide can be expressed as
\begin{align} \label{basic_channel_model}
    h 
    = \underbrace{\frac{\eta \exp\left(-j \tfrac{2\pi}{\lambda} R\right)}{R}}_{\text{free-space propagation}}
      \times
      \underbrace{\vphantom{\frac{\eta e^{-j \beta_0 r}}{r}} \rho(\theta,\phi)\, \kappa}_{\text{radiation \& coupling}}
      \times
      \underbrace{\vphantom{\frac{\eta e^{-j \beta_0 r}}{r}} \exp\left(-j \frac{2\pi}{\lambda_g} L\right)}_{\text{in-waveguide propagation}}.
\end{align}
In this model, $\lambda$ is the free-space wavelength, and $\eta = \lambda / (4\pi)$ captures the wavelength-dependent amplitude factor of the free-space Green’s function. The term $\rho(\theta,\phi)$ denotes the PA radiation pattern in the direction $(\theta,\phi)$, and $\kappa$ is the coupling factor that determines the fraction of guided power extracted and radiated by the PA. Both $\rho(\theta,\phi)$ and $\kappa$ depend on the specific hardware implementation of the PA. For analytical studies, it is often convenient to consider an ideal isotropic PA as a reference design, for which $\rho(\theta,\phi) = 1, \forall\, \theta, \phi$, so that the angular dependence is removed and the impact of waveguide propagation and coupling can be isolated more transparently.

In architectures where multiple PAs are attached to a single waveguide, the signal propagation is inherently cascaded, requiring a model that accounts for sequential power extraction. Consider a system comprising $N$ PAs, where the geometry for the $n$-th PA is defined by the in-waveguide propagation distance $L_n$ and the free-space propagation distance to the user $R_n$. A critical factor in this serial arrangement is the coupling mechanism. Let $\kappa_n$ denote the coupling factor and $\rho_n(\theta, \phi)$ denote the radiation pattern for the $n$-th PA. Assuming matched ports with negligible reflection, the signal amplitude is progressively attenuated as it passes each PA. Specifically, the transmission coefficient past the $i$-th PA is $\sqrt{1 - \kappa_i^2}$, representing the signal remaining in the waveguide. Consequently, the effective, cascaded coupling factor for the $n$-th PA is the product of its own coupling and the cumulative transmission of all preceding PAs, given by $\widetilde{\kappa}_n = \kappa_n \prod_{i=1}^{n-1} \sqrt{1 - \kappa_i^2}$. By jointly considering the free-space path loss, the directional radiation pattern, this effective coupling factor, and the in-waveguide phase shift, the end-to-end baseband channel response $h_n$ for the $n$-th PA is formulated as
\begin{align}
    h_n 
    = \frac{\eta \exp\left(-j \tfrac{2\pi}{\lambda} R_n\right)}{R_n}
      \times
      \rho_n(\theta_n,\phi_n)\, \widetilde{\kappa}_n
      \times
       \exp\left(-j \frac{2\pi}{\lambda_g} L_n\right),
\end{align}     
where $(\theta_n, \phi_n)$ represents the user's direction relative to the $n$-th PA. The total signal received at the user is the superposition of these individual channel contributions plus additive noise $w(t)$, expressed as follows:
\begin{align}
    y(t) = \left(\sum_{n=1}^N h_n \right) s_{\mathrm{in}}(t) + w(t).
\end{align}  
From a system design perspective, it is crucial to observe that both the free-space distance $R_n$ and the in-waveguide distance $L_n$ are functions of the physical positions of the PAs. This geometric dependence provides a unique degree of freedom: By carefully designing the PA spacing, one can align the phases of the individual channel components $h_n$ to maximize the coherent summation. This optimization strategy yields a strong overall channel gain $\sum_{n=1}^N h_n$ and is referred to as \emph{pinching beamforming}, which will be elaborated upon in the following sections.

\subsection{PA Implementation and Hardware Models}
As established in the signal and channel model, the PA plays a critical role in PASS by extracting signals from the waveguide and radiating them into free space. This functionality is captured by the radiation and coupling factors, $\rho(\theta, \phi)$ and $\kappa$, in Eq.~\eqref{basic_channel_model}. These factors are grounded in EM theory and are determined by the specific hardware implementation of the PAs.

\begin{figure}[t]
    \centering
    \subfigure[Rectangular Waveguide.]{
        \includegraphics[width=0.43\linewidth]{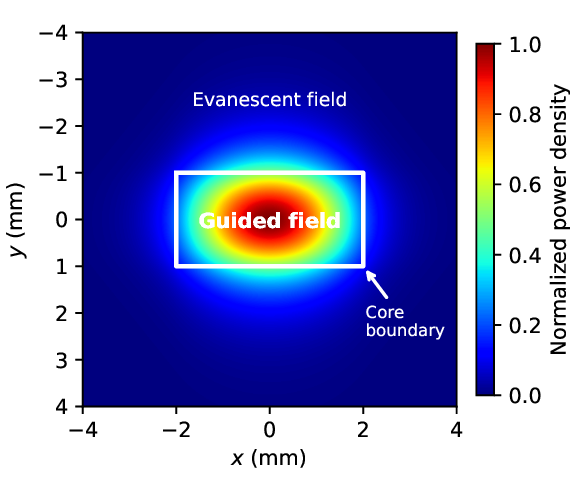}
        \label{fig:rectangular_guide}
    }
    \subfigure[Circular Waveguide]{
        \includegraphics[width=0.43\linewidth]{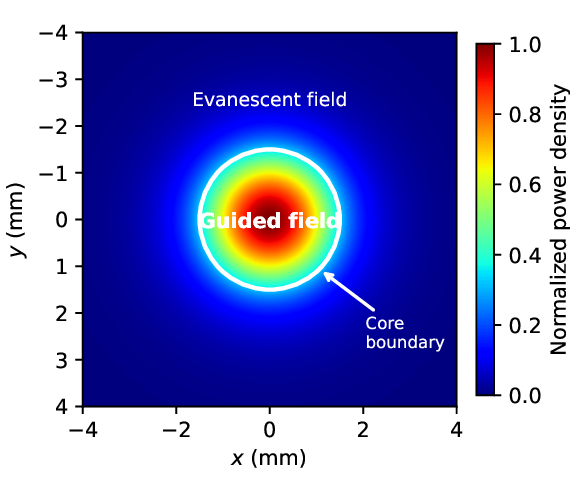}
        \label{fig:circular_guide}
    }
    \caption{Illustration of the evanescent and propagation fields for rectangular and circular waveguides. The figures illustrate the power density at 28 GHz for a dielectric core with a refractive index of 2.1 surrounded by air. Only the fundamental mode is plotted.}
    \label{fig:evanescent_field}
\end{figure}

\begin{figure*}[t]
    \centering
    \subfigure[SDS Model.]{
        \includegraphics[width=0.33\linewidth]{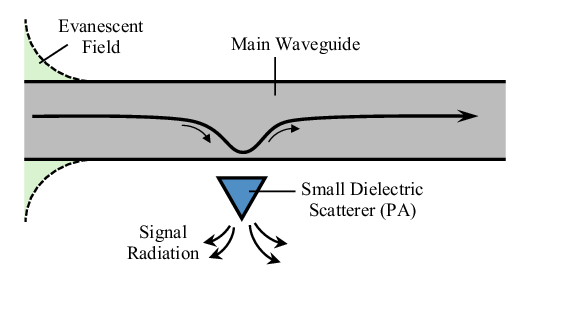}
        \label{fig:SDS}
    }
    \hspace{-0.6cm}
    \subfigure[DCW Model -- Leaky Mode.]{
        \includegraphics[width=0.33\linewidth]{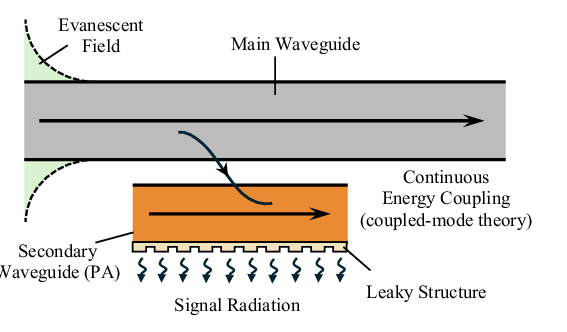}
        \label{fig:DCW_leaky}
    }
    \hspace{-0.6cm}
    \subfigure[DCW Model -- Feeder Mode.]{
        \includegraphics[width=0.33\linewidth]{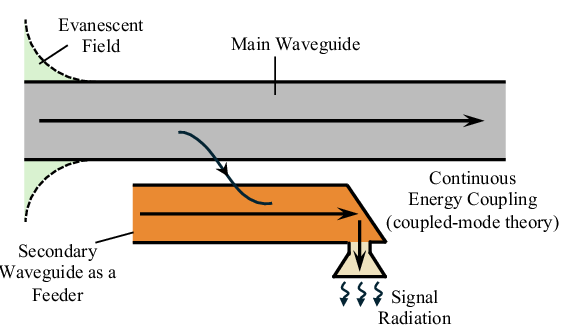}
        \label{fig:DCW_feeder}
    }
    \caption{Three EM-level hardware implementations for PAs.}
    \label{fig:hardware_model}
\end{figure*}

The foundational physical principle of a PA relies on disrupting the normal transmission properties of a dielectric waveguide to intentionally leak radio waves at specific controllable locations. For a waveguide, while the propagating fields are confined within its core, the evanescent fields extend beyond its boundary \cite{Marcuse1974}, as illustrated in Fig. \ref{fig:evanescent_field}. Consequently, when a secondary dielectric element is brought into the evanescent field region (e.g., via pinching), it extracts a portion of the wave propagating along the main waveguide. This coupled wave is then radiated outward, effectively transforming the secondary element into an antenna. This physical phenomenon is entirely reversible. Specifically, if the PA, i.e., the secondary dielectric element, is removed, the radiation ceases, and the main waveguide returns to acting solely as a transmission line. These unique features introduce significant flexibility. On the one hand, radiation points can be selected as desired by simply moving the pinching point. On the other hand, the antenna is formed or eliminated simply by attaching or removing the secondary dielectric element. 

The radiation and coupling factors of a PA rely on the interaction between the secondary dielectric element and the evanescent field surrounding the main waveguide. The behavior of this interaction changes depending on the physical scale and geometry of the PA, leading to different hardware implementation models, as elaborated below.

\subsubsection{Small Dielectric Scatterer (SDS) Model} 
In this model, the PA operates as a small dielectric scatterer introduced into the proximity of the main waveguide, as depicted in Fig. \ref{fig:SDS}, closely matching the NTT DOCOMO prototype~\cite{suzuki2022pinching}. As mentioned before, the waveguide confines high-frequency waves within its core via total internal reflection. However, the evanescent field decays exponentially outside the core boundaries. When the secondary dielectric element is present in this region, it perturbs and scatters the evanescent field. Following the Huygens' equivalence principle \cite{237620}, this small scatterer effectively acts as a source of equivalent polarization currents induced by the evanescent field and radiates energy into free space. Because the scatterer is electrically small and acts as a localized perturbation, the propagation mode in the main waveguide remains nearly unchanged. Based on the above principle, the communication zone can be dynamically established or removed by simply attaching or detaching the small scatterer.

From an implementation perspective, the SDS model offers significant mechanical simplicity by eliminating the need for complex fabrication. For example, a simple dielectric block in the shape of a triangular pyramid can be pinched onto the main waveguide to form a PA, allowing for rapid and cost-effective deployment. However, this approach has low power efficiency and limited controllability. Since an electrically small scatterer interacts only with a small fraction of the evanescent field, the coupling coefficient is inherently low. Furthermore, this model lacks a closed-form analytical theory that quantitatively correlates the in-waveguide power to the radiated power. Without this theory, it is difficult to adjust the physical geometry of PAs to achieve specific radiation properties, such as a particular radiation pattern or power coupling ratio.

\begin{table*}[t]
\centering
\caption{Summary of different models of PAs.}
\label{table:PA_models}
\resizebox{1\textwidth}{!}{
\begin{tabular}{|llc|l|l|l|l}
\cline{1-6}
\multicolumn{3}{|c|}{\textbf{Models}}                                                                                                                             & \textbf{Physics} \textbf{Basis}                                       & \textbf{Key Pros}                                                                                                               & \textbf{Key Cons}                                                                  &  \\ \cline{1-6}
\multicolumn{1}{|l|}{\multirow{4}{*}{\makecell{EM-level\\ Hardware Model}}} & \multicolumn{2}{c|}{SDS}                                   & \makecell[l]{Perturbation theory,\\Huygens’ equivalence principle} & \makecell[l]{Simple fabrication,\\easy to attach/detach}                                                       & \makecell[l]{Low power efficiency, \\ low controllability}                   &  \\ \cline{2-6}
\multicolumn{1}{|l|}{}                                         & \multicolumn{1}{l|}{\multirow{3}{*}{DCW}} & Leaky mode  & \makecell[l]{Coupled-mode theory, \\ leaky wave structure}          & \makecell[l]{Stronger coupling than SDS, \\ integrated design}                                               & \makecell[l]{Lower controllability\\ than Feeder mode}                               &  \\ \cline{3-6}
\multicolumn{1}{|l|}{}                                         & \multicolumn{1}{l|}{}                                                     & Feeder mode & \makecell[l]{Coupled-mode theory,\\dedicated radiator}            & \makecell[l]{Decoupled coupling and radiation, \\high-gain beams}                                                             & \makecell[l]{Complex fabrication, \\ tight mechanical tolerances}                &  \\ \cline{1-6}
\multicolumn{3}{|l|}{\makecell{Multiport Network Model}}                                                                                               & \makecell[l]{Multiport theory,\\circuit theory}                    & \begin{tabular}[c]{@{}l@{}}\makecell[l]{Simple system abstraction, \\ hardware imperfection consideration}\end{tabular} & Limited physical insight &  \\ \cline{1-6}
\end{tabular}
}
\end{table*}

\subsubsection{Directional Coupler Waveguide (DCW) Model} 

To enhance the radiation efficiency and controllability, the secondary dielectric element can be elongated into a strip rather than a point. In this case, it can be modeled as a secondary short waveguide running parallel to the main waveguide, as illustrated in Fig. \ref{fig:DCW_leaky} and Fig. \ref{fig:DCW_feeder}. The interaction between the main waveguide and the secondary element is described by coupled-mode theory \cite{Marcuse1974, Miller1954}, where energy continuously tunnels between them according to their coupling strength. Compared with the SDS model, the DCW model enables stronger and more controllable power extraction from the main waveguide. To radiate the coupled energy into free space, two operational modes can be employed:
\begin{itemize}
    \item \textbf{The Leaky Mode:} As shown in Fig. \ref{fig:DCW_leaky}, in this mode, the secondary dielectric element behaves similarly to a short LCX. As energy couples from the main waveguide into the secondary element, the latter is designed to simultaneously leak this energy into the air through its tailored material properties or geometry. In the leaky-mode DCW model, the secondary element itself serves as the radiator, where coupling and radiation occur concurrently, forming a complete PA. The radiation from the start to the end of the coupling region adds constructively in the forward direction, resulting in a directional bias.
    \item \textbf{The Feeder Mode:} As shown in Fig. \ref{fig:DCW_feeder}, in this mode, the secondary dielectric element acts primarily as a feeder rather than as a direct radiator. It is designed to efficiently extract energy from the main waveguide with minimal leakage along its length. Once the energy is transferred into the secondary element, it is guided to the element's terminus, where it connects to a dedicated radiator (e.g., through a 45$^\circ$ mirror \cite{morimoto2019design, 9103289}) to form the complete PA assembly. This architecture effectively decouples the power coupling from the radiation control, enabling the design of high-gain directional beams. The feeder-mode DCW represents a high-performance implementation method, but its high efficiency comes at the cost of substantial implementation complexity and tight mechanical tolerances. In particular, the interface between the secondary dielectric element and the dedicated radiator must be carefully impedance-matched to suppress signal reflections.
\end{itemize}

\begin{figure}[t]
    \centering
    \includegraphics[width=0.45\textwidth]
    {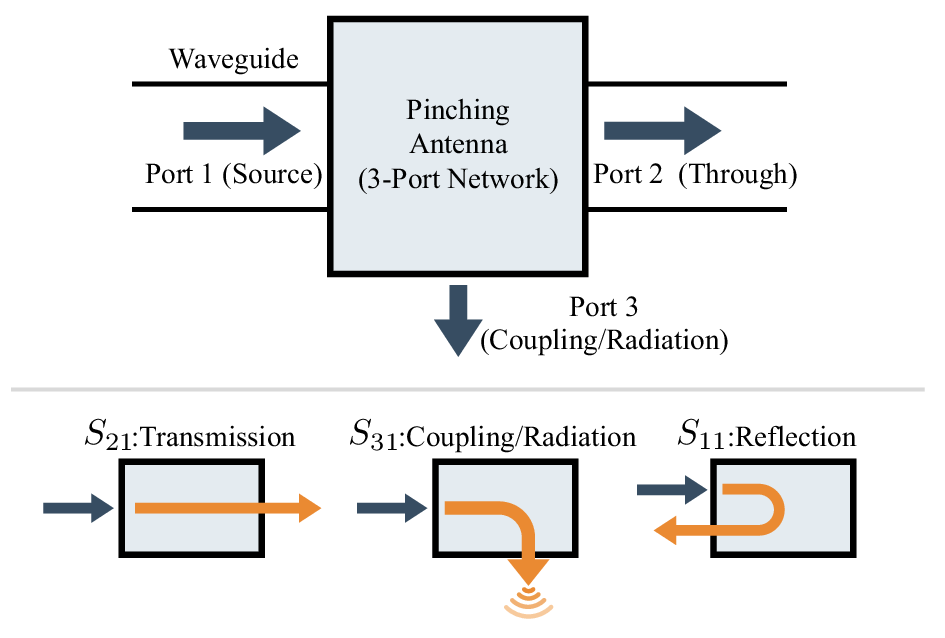}
    \caption{Multiport network model for PAs.}
    \label{fig:multiport}
\end{figure}

The DCW model, compared to the SDS model, enables higher radiation efficiency and controllability, thus attracting significant attention. The DCW model was first proposed for PAs in early work \cite{wang2025modeling}, where the power coupling factor was characterized using coupled-mode theory. Subsequently, equal-power and proportional-power models were proposed for multiple PAs on the same waveguide. The model was then extended to wideband orthogonal frequency division multiplexing (OFDM) systems in \cite{xiao2025freqeuncy}, where the frequency-dependent coupling factor and waveguide dispersion were examined. Furthermore, the authors of \cite{xu2025pinching} developed a more flexible model with an adjustable coupling factor. They achieved amplitude-only control of the radiated signal by revealing the relationship between the coupling factor and the gap between the PA and the waveguide. Another advance was the conception of a circuit-based DCW model in \cite{wang2025multiport}, which unveiled mechanisms for amplitude-only and amplitude-constrained phase control. Finally, the authors of \cite{zhang2025directional} analyzed the directionality of the radiation pattern in the leaky-mode DCW model and characterized the optimal orientation of the PAs.

\subsection{PA Multiport Network Model}

The previous section introduced three implementation methods and corresponding theories for their analysis. While these theories are useful for predicting and analyzing PA behavior, they are often insufficient for accurately quantifying key parameters such as radiation and coupling efficiency. Furthermore, it remains difficult to account for hardware imperfections, such as signal reflections caused by impedance mismatches at pinching points. Given the challenges in accurately characterizing the complex EM behavior of these implementations, multiport network theory provides a useful macroscopic abstraction. This theory models the PA as a lumped three-port network, as illustrated in Fig. \ref{fig:multiport}. In this model, the main waveguide functions as a transmission line connecting Port 1 (Source) and Port 2 (Through), while the PA introduces a Port 3, representing the radiation and coupling mechanism \cite{Pozar2011, wang2025multiport}. By employing this theoretical framework, system behavior is characterized not by solving complex field integrals, but by analyzing the scattering parameters shown in Fig. \ref{fig:multiport}. The physical interpretations of these parameters are detailed below:
\begin{itemize}
    \item \textbf{Signal Transmission ($S_{21}$):} Quantifies the signal remaining in the main waveguide after passing the PA. A reduction in magnitude here reflects the power extracted for radiation.
    \item \textbf{Radiation and Coupling ($S_{31}$):} Represents the radiation and coupling efficiency from the main waveguide to the PA. It defines the fraction of guided power successfully converted into radiated emissions. By relating this to Eq. \eqref{basic_channel_model}, we obtain $S_{31} = \rho(\theta, \phi) \kappa$ under reflection-free conditions.
    \item \textbf{Reflection ($S_{11}$):} Measures the signal reflected back toward the source. This arises from the local impedance mismatch introduced by the presence of the PA. In most cases, minimizing this reflection is critical for ensuring system stability and simplifying impedance matching requirements.
\end{itemize}

The multiport network model serves as a powerful tool for system-level analysis by treating the PA as a lumped three-port network. Its primary advantage lies in its simplicity, as it abstracts complex EM interactions into straightforward scattering parameters, which can be measured in practice or simulated via full-wave simulation. Based on the multiport network model, the authors of \cite{wang2025multiport} provided a comprehensive study of the PASS signal model for both single-PA and multiple-PA deployments. However, this abstraction comes at the cost of physical insight. Specifically, the model provides no information regarding physical characteristics such as coupling strength, radiation pattern, and polarization, providing limited guidance to the designer.

Table \ref{table:PA_models} provides a comparison between different PA models discussed above.

\subsection{Discussion and Outlook}

In this section, we established the fundamentals of PASS, moving from signal and channel modeling to physical insights. % However, several key issues remain open. 
Although DOCOMO's prototype provides an important proof-of-concept for PASS, it does not constitute a complete engineering validation for large-scale deployment. Practical validation should further connect hardware characterization, calibrated channel modeling, and system-level multi-user evaluation.
In the following paragraphs, we elaborate on three major points.

\subsubsection{Hardware Limitations}
The SDS model is mechanically simple and easy to deploy, but its coupling efficiency is usually low and difficult to calibrate because the scattered field is highly sensitive to the shape, size, material, and contact condition of the dielectric scatterer. The DCW model improves power extraction and controllability, but it requires accurate control of the coupling length, waveguide gap, alignment, material loss, and termination or radiator matching. In addition, PA reconfiguration involves physical movement rather than purely electronic switching. Therefore, actuator latency, positioning error, mechanical wear, and mounting tolerances may limit the ability of PASS to track fast user movement. Moreover, dielectric constants, loss tangents, and coupling coefficients are frequency-dependent, so a PA design optimized at one carrier frequency may not directly support wideband or multi-band operation.

\subsubsection{System Design Simplifications}
Many system-level PASS models assume unidirectional in-waveguide propagation, matched ports, negligible reflection, and independent PA radiation. These assumptions are useful for obtaining tractable signal models, but they may be inaccurate in practical deployments. Each PA perturbs the local impedance of the waveguide and can generate backward reflections, which are characterized by the reflection coefficient $S_{11}$ in the multiport network model. Closely spaced PAs may also experience mutual coupling and inter-PA radiation, thereby changing the effective transmission coefficient $S_{21}$ and radiation coefficient $S_{31}$ of neighboring PAs. Consequently, the radiated power ratios, phases, and radiation patterns can become calibration-dependent. Reliable hardware-aware modeling therefore requires full-wave simulation, measurement-based calibration, or multiport network characterization.

The modeling of propagation delay is another important issue for future wideband or OFDM-integrated PASS designs. 
For the narrowband model, the propagation delay is typically absorbed into the in-waveguide phase rotation at a specific carrier frequency. 
For wideband or OFDM-integrated PASS, however, the frequency-dependent propagation constant may introduce non-negligible waveguide dispersion, and explicit modeling of propagation delays and delay spread should be considered.

\subsubsection{Optimization Limitations}
PASS optimization is also constrained by these hardware realities. Unlike conventional antenna arrays, where complex weights can usually be adjusted electronically, PASS often relies on the physical positions and coupling states of PAs. These variables are jointly coupled through in-waveguide propagation, free-space propagation, reflection, mutual coupling, actuator constraints, and allowable deployment locations. As a result, the resulting optimization problem is highly non-convex and cannot be fully captured by ideal continuous-position models. Practical algorithms should incorporate finite positioning resolution, minimum PA spacing, available coupling ranges, calibration errors, and reconfiguration time in addition to conventional communication performance metrics. Detailed discussions about promising optimization methods are provided in Section \ref{sec:optimization}.

\section{Promising PASS Architectures With Scalable DoFs} \label{sect:promising}
Having established the fundamental principles of PASS together with its practical implementation methods, we next introduce several promising design architectures for PASS that build upon the implementation approaches discussed in the previous section. In particular, three structures are highlighted: segmented PASS, center-fed PASS, and multi-mode PASS.
\subsection{Segmented Waveguide-Enabled Pinching-Antenna System (SWAN)}
\subsubsection{Motivation of SWAN}

\begin{figure}[!t]
\centering
    \subfigure[System setup.]
    {
        \includegraphics[width=0.45\textwidth]{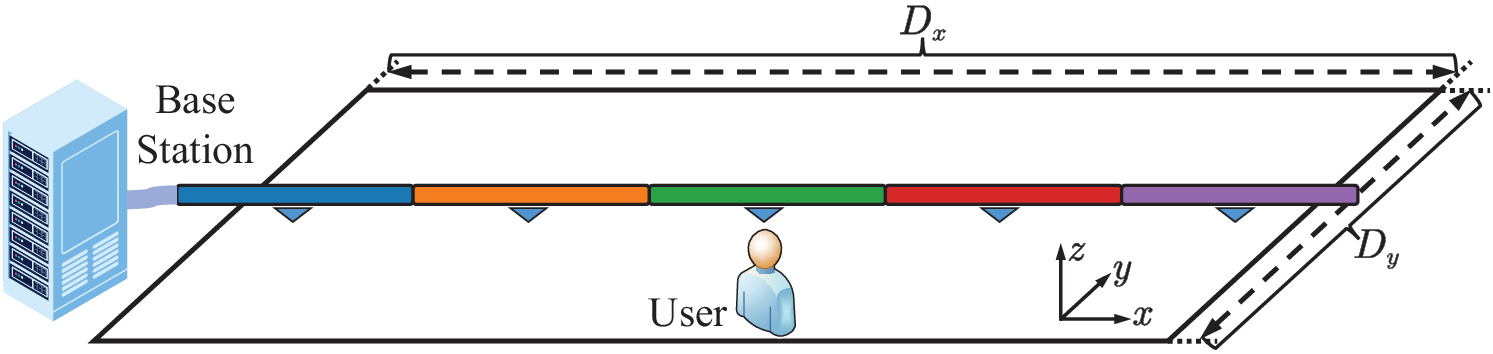}
	   \label{Figure: PAS_System_Model1}
    }
   \subfigure[Segmented waveguide.]
    {
        \includegraphics[width=0.45\textwidth]{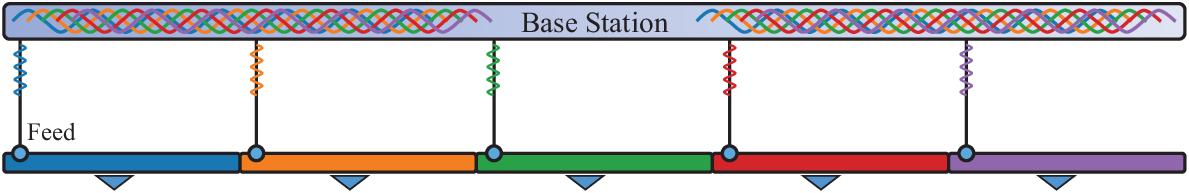}
	   \label{Figure: PAS_System_Model2}
    }
\caption{Illustration of the proposed SWAN architecture.}
\label{Figure: PAS_System_Model}
\vspace{-15pt}
\end{figure}

\begin{figure*}[!t]
\centering
    \subfigure[Segment selection.]
    {
        \includegraphics[height=0.15\textwidth]{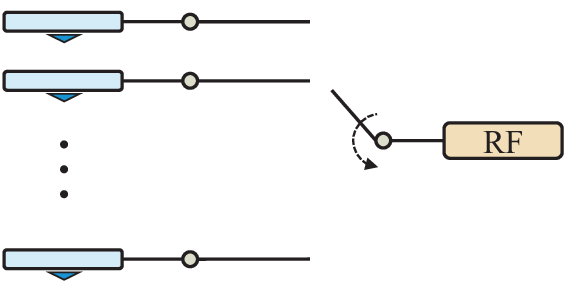}
	   \label{Figure_PAN_Protocol1}
    }
   \subfigure[Segment aggregation.]
    {
        \includegraphics[height=0.15\textwidth]{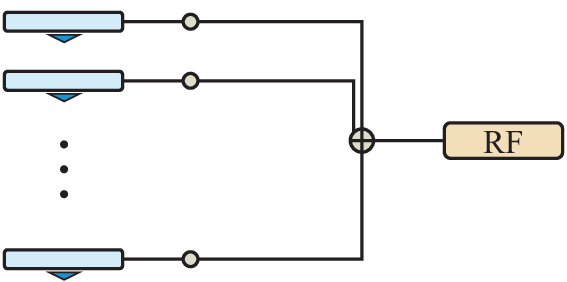}
	   \label{Figure_PAN_Protocol2}
    }
    \subfigure[Segment multiplexing.]
    {
        \includegraphics[height=0.15\textwidth]{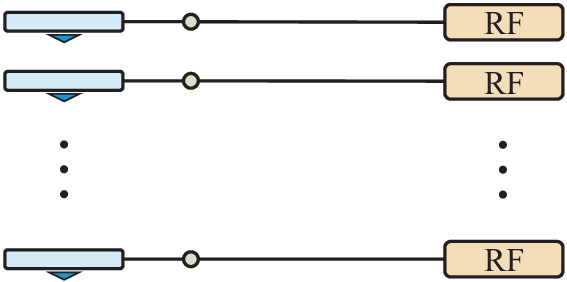}
	   \label{Figure_PAN_Protocol3}
    }
\caption{Illustration of three basic protocols for operating SWANs.}
\label{Figure: PAN_Protocol}
\end{figure*}

Despite significant research progress in PASS, existing architectures face three key challenges.
\begin{itemize}
  \item Uplink Signal Model: The downlink signal model of PASS has been clearly characterized, where signals injected at the waveguide feed point radiate passively from the activated PAs based on the EM coupling model proposed in \cite{wang2025modeling,liu2025pinching}. In contrast, formulating a tractable and physically consistent uplink signal model is much more challenging. This difficulty arises because PAs can passively receive EM signals from free space into the waveguide. Furthermore, when multiple PAs are deployed along the same waveguide, signals captured by one PA may re-radiate through other PAs as they propagate toward the feed point. This \emph{inter-antenna radiation (IAR)} effect complicates the uplink analysis and makes the signal model mathematically intractable. As a result, most existing uplink studies restrict attention to single-PA deployments, thereby avoiding IAR, or they neglect IAR altogether in multi-PA systems, which leads to oversimplified models \cite{liu2025pinching}. At present, no tractable and physically consistent uplink signal model for multi-PA systems exists, leaving research in this direction stalled.
  \item In-Waveguide Propagation Loss: Another challenge arises when a single long waveguide is deployed to cover a large region. While flexible PA placement can reduce the average user-to-antenna distance and thereby alleviate large-scale path loss, a long waveguide simultaneously increases the propagation distance between PAs and the feed point. This, in turn, increases the in-waveguide propagation loss. Such loss is negligible for small or moderate waveguide lengths, but becomes a major limiting factor in large-scale deployments. The numerical results in this work demonstrate that in-waveguide loss can significantly degrade system performance when long waveguides are employed.
  \item Waveguide Maintainability: Finally, long waveguide deployments present practical challenges in terms of reliability and maintenance. If a waveguide is damaged at a particular point, fault detection may be difficult, and replacing the entire waveguide can be both costly and complex. This lack of robustness reduces the practical feasibility of long-waveguide PASS deployments, as they are fragile and difficult to repair once installed.
\end{itemize}
To address the aforementioned challenges, the \emph{Segmented Waveguide-enabled pinching-antenna system (SWAN)} has been proposed \cite{ouyang2025uplink}. The SWAN architecture employs multiple short dielectric-waveguide segments arranged end-to-end. Unlike a single long waveguide, these segments are \emph{not physically interconnected}. Instead, each segment has its own feed point, through which signals are injected into or extracted from the waveguide and then relayed to the base station (BS) via wired connections such as optical fiber or high-quality coaxial cables, as illustrated in {\figurename} {{\ref{Figure: PAS_System_Model}}}. 

The proposed SWAN overcomes the limitations of conventional PASS in three ways. First, SWAN activates only a single PA within each segment, enabling the realization of a multi-PA uplink PASS without IAR effects. This approach allows the system to achieve a multi-PA array gain while maintaining a tractable uplink signal model. Second, since each segment is much shorter than the overall service region, SWAN ensures a reduced average PA-to-user distance and shortens the PA-to-feed distance. This dual benefit mitigates large-scale path loss and in-waveguide propagation loss. Third, the segmented design enhances maintainability since failures can be localized and addressed at the segment level, rather than requiring the replacement of the entire waveguide. Damaged short segments are easier to detect, repair, and replace than a damaged single long waveguide. Finally, while long dielectric waveguides can be fabricated, existing PASS prototypes developed by NTT DOCOMO use short waveguides ranging from 0.4 to 1.2 meters in length \cite{yamamoto2021pinching,reishi2022pinching,junya2024pinching,junya2025pinching}. Thus, the SWAN architecture can be readily implemented by leveraging existing PASS hardware with only minimal architectural modifications.
\subsubsection{Operating Protocols of SWAN}
Referring to {\figurename} {\ref{Figure: PAS_System_Model2}}, the performance of the SWAN depends on the connection mechanism between the feed points and the RF front-end of the BS. Motivated by this observation, we propose three basic operating protocols: \romannumeral1) \emph{segment selection (SS)}, \romannumeral2) \emph{segment aggregation (SA)}, and \romannumeral3) \emph{segment multiplexing (SM)}, as illustrated in {\figurename} {\ref{Figure: PAN_Protocol}}.
\begin{itemize}
  \item For SS, as shown in {\figurename} {\ref{Figure_PAN_Protocol1}}, only a single selected segment is connected to the RF chain at a given time. This protocol is straightforward to implement using a simple switching mechanism and incurs very low hardware complexity.
  \item For SA, as illustrated in {\figurename} {\ref{Figure_PAN_Protocol2}}, all feed points are connected to a single RF chain via a power splitter. In the uplink, signals extracted from all segments are aggregated and forwarded to the RF chain for baseband processing. In the downlink, the transmit signal is equally split across the segments and then radiated to the user through the PAs attached to them. SA is expected to outperform SS, as all waveguide segments contribute simultaneously, though it requires a more complex RF front-end to enable power splitting and aggregation. 
  \item For SM, as illustrated in {\figurename} {\ref{Figure_PAN_Protocol3}}, each waveguide segment is connected to its own dedicated RF chain. This setup enables optimal digital processing, such as maximal-ratio combining (MRC) in the uplink and maximal-ratio transmission (MRT) in the downlink, thereby achieving the performance upper bound of the SWAN. However, SM entails significantly higher hardware complexity than SS and SA due to the requirement of multiple RF chains and advanced baseband processing.
\end{itemize}

\begin{table*}[!t]
\centering
\caption{Summary of the Key Features of the Considered Operating Protocols}
\setlength{\abovecaptionskip}{0pt}
\resizebox{0.98\textwidth}{!}{
\begin{tabular}{|l|l|l|l|l|l|}
\hline
\textbf{Protocol} & \textbf{Architecture}                                  & \textbf{RF Chain} & \textbf{Performance} & \textbf{Implementation Complexity}& \textbf{Optimization Variables} \\ \hline
\textbf{Segment Selection}       & Only one segment connects to the RF chain     & 1        & Lowest      & Very low (switch)  & Antenna positions and activated segment                \\ \hline
\textbf{Segment Aggregation}       & All segments are aggregated into one RF chain & 1        & Moderate    & Moderate (power splitter)   & Antenna positions        \\ \hline
\textbf{Segment Multiplexing}       & Each segment has its own RF chain             & $M$      & Highest     & High (multi-RF chain hardware) & Antenna positions and baseband beamforming                     \\ \hline
\end{tabular}}
\label{Table: PASS_Protocol_Comparision}
\end{table*}

Table \ref{Table: PASS_Protocol_Comparision} provides a summary of the key characteristics of the SS, SA, and SM protocols.
\subsubsection{Tri-Hybrid Beamforming Structures}
\begin{figure}[!t]
\centering
\includegraphics[height=0.12\textwidth]{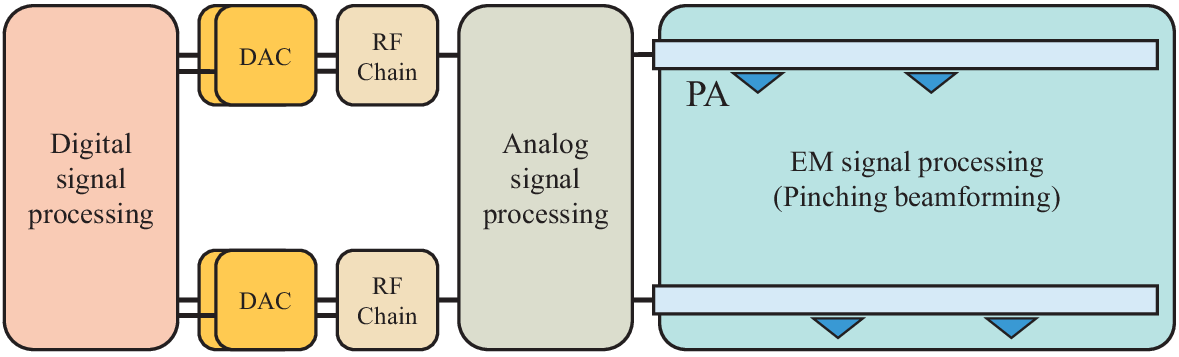}
\caption{Illustration of the tri-hybrid beamforming architecture.}
\label{System_SWAN_Tri}
\end{figure}

Beyond the three operating protocols, SWAN can also support a \emph{tri-hybrid beamforming architecture}, as shown in {\figurename} {\ref{System_SWAN_Tri}}. This structure integrates three layers of spatial processing: \emph{Digital beamforming}, \emph{analog beamforming}, and \emph{EM beamforming} enabled by pinching antennas. The first two layers follow the conventional hybrid MIMO framework, while the third layer, EM beamforming via pinching antennas, introduces an additional degree of spatial control \cite{castlanos2025embracing,heath2025tri}. This extra layer allows the system to adjust the effective aperture and radiation characteristics at the EM level, enhancing design flexibility without increasing the number of RF chains or power consumption.

\begin{figure}[!t]
\centering
    \subfigure[Phase shifter-based.]
    {
        \includegraphics[height=0.11\textwidth]{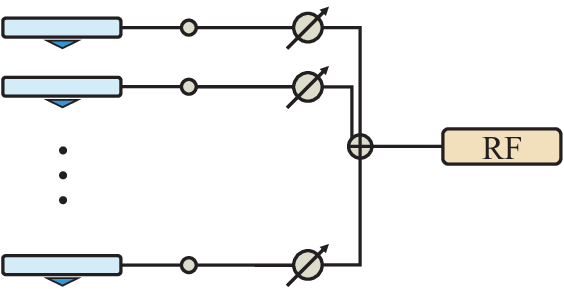}
	   \label{Figure_PAN_Protocol5}
    }
    \subfigure[Switch-based.]
    {
    \includegraphics[height=0.11\textwidth]{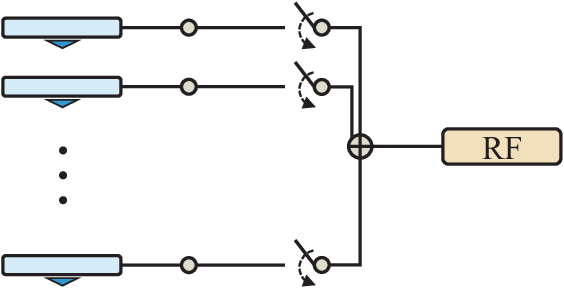} \label{Figure_PAN_Protocol4}
    }
\caption{Illustration of two extended protocols for SAs.}
\label{Figure: PAN_Protocol_Ex}
\end{figure}

We begin with the special case where all waveguide segments are connected to a single RF chain, which can be viewed as an extension of the SS and SA protocols. In this setting, two tri-hybrid beamforming architectures, one based on phase shifters and the other on switching networks, can be designed, as illustrated in {\figurename} {\ref{Figure: PAN_Protocol_Ex}}. In the phase-shifter-based architecture, each segment is connected to the RF chain through a digitally controlled phase shifter with a small number of quantized phase states. This enables an additional layer of analog beamforming across the segments, providing further flexibility to enhance system performance. Alternatively, a switching-based tri-hybrid architecture can be adopted, where low-loss RF switches are used instead of phase shifters. This reduces hardware complexity and power consumption. This design is motivated by the analytical results in \cite{ouyang2025uplink}, which indicate that activating all segments does not always maximize the received signal-to-noise ratio (SNR), and there exists an optimal number of active segments. Thus, the switching network allows the system to select only the most effective subset of segments. By integrating these two structures, a dynamic tri-hybrid beamforming architecture can be further developed, where switches first select the subset of active segments, and phase shifters then refine their analog beamforming weights. This hybrid mechanism provides flexibility to balance performance and hardware cost.
As an example, a novel SWAN structure is proposed and validated in \cite{gu2026joint}.

Since all of the above architectures rely only on one RF chain, they can be readily combined with practical multiple access protocols such as time-division multiple access (TDMA), orthogonal frequency-division multiple access (OFDMA), and non-orthogonal multiple access (NOMA) to support multiple users. For orthogonal multiple access (OMA), two strategies may be considered:
\begin{itemize}
  \item \textbf{\emph{PA Switching}}: The PAs' positions are optimized separately for each user within its allocated resource block.
  \item \emph{\textbf{PA Multiplexing}}: The PAs' positions are jointly optimized for all users served simultaneously. 
\end{itemize}

\begin{figure}[!t]
\centering
    \subfigure[Fully-connected.]
    {
        \includegraphics[height=0.12\textwidth]{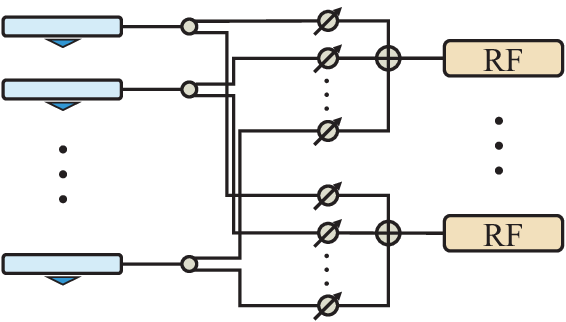}
	   \label{Figure_PAN_Tri_Phase_Full}
    }
   \subfigure[Sub-connected.]
    {
        \includegraphics[height=0.12\textwidth]{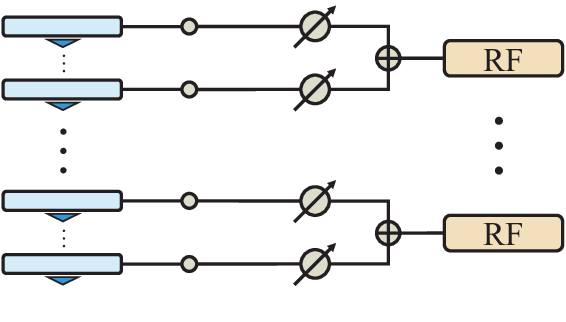}
	   \label{Figure_PAN_Tri_Phase_Sub}
    }
\caption{Analog and EM processing for tri-hybrid beamforming based on phase shifters: (a) each RF chain can be connected to all the segments; (b) each RF chain can be connected to a subset of segments.}
\label{Figure_PAN_Tri_Phase}
\end{figure}

\begin{figure}[!t]
\centering
    \subfigure[Fully-connected.]
    {
        \includegraphics[height=0.11\textwidth]{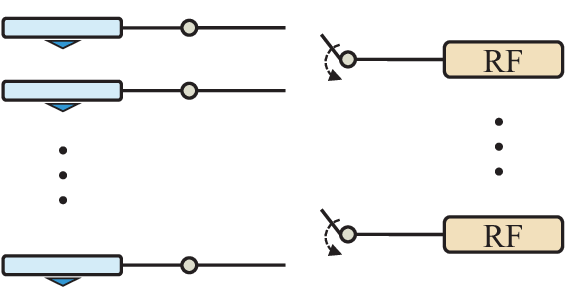}
	   \label{Figure_PAN_Tri_Switch_Full}
    }
   \subfigure[Sub-connected.]
    {
        \includegraphics[height=0.11\textwidth]{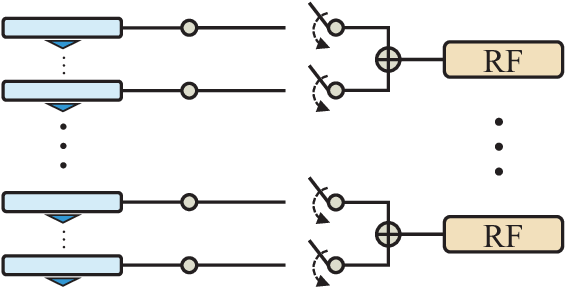}
	   \label{Figure_PAN_Tri_Switch_Sub}
    }
\caption{Analog and EM processing for tri-hybrid beamforming based on switches: (a) each RF chain can be connected to all the segments; (b) each RF chain can be connected to a subset of segments.}
\label{Figure_PAN_Tri_Switch}
\end{figure}

We next consider the case where multiple RF chains are available at baseband to enable digital beamforming. Two realizations of the tri-hybrid architecture are illustrated in {\figurename} {\ref{Figure_PAN_Tri_Phase}}. In the first structure ({\figurename} {\ref{Figure_PAN_Tri_Phase_Full}}), all waveguide segments are connected to every RF chain, thereby enabling maximum spatial flexibility but at the cost of higher hardware complexity. In the second structure ({\figurename} {\ref{Figure_PAN_Tri_Phase_Sub}}), the segment-array is partitioned into multiple sub-segment-arrays, where each sub-segment-array is connected to an individual RF chain. Although this reduces hardware complexity, it provides lower spatial configurability compared to the full-connection architecture. In both architectures, the analog beamforming stage can be implemented with digitally controlled phase shifters that have a finite set of quantized phase states. Then, the digital precoder/combiner can refine the beamforming weights to compensate for imperfections in the analog and EM beamforming layers. For example, it can suppress residual multi-stream interference.

Alternatively, low-loss switching networks can replace phase shifters, further reducing system complexity and power consumption. As shown in {\figurename} {\ref{Figure_PAN_Tri_Switch}}, the switching-based design leverages the sparse nature of high-frequency propagation by performing compressed spatial sampling of the received waveform. In this case, the analog combiner is designed by selecting an optimized subset of waveguide segments rather than optimizing all quantized phase values. Each switch can connect to all segments when the array is small or to a subset of segments when the array is large.

Together, these four tri-hybrid architectures support the development of multiuser beamforming algorithms that jointly optimize digital, analog, and EM pinching beamforming to take full advantage of the spatial DoFs enabled by SWAN.

\subsection{Center-Fed PASS (C-PASS)}
\begin{figure*}[t]
	\centering
    \includegraphics[width=0.9\linewidth]{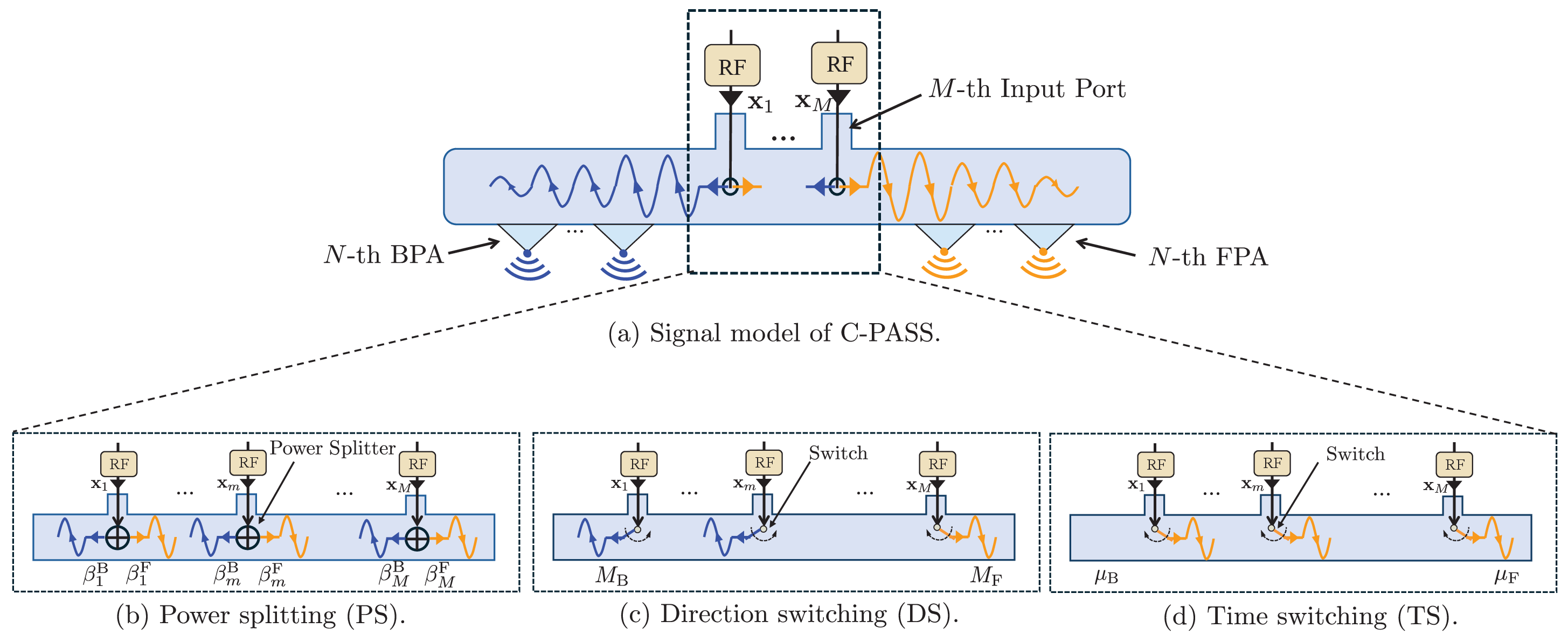}
	\caption{Three practical protocols for operating C-PASS.}
	\label{fig:C-PASS}
\end{figure*}

\begin{table}[t]
	\centering
    \small
    \caption{Comparison between center-fed and end-fed PASS.}
	\label{tab:analysis_CPASS}
		\begin{tabular}{|l|c|c|c|} 
			\hline 
			\textbf{Architecture} & \textbf{DoF} & \textbf{Array Gain} & \textbf{Multiplexing Gain} \\
			\hline 
			Center-Fed PASS & 2 & $\mathcal{O}\left( \frac{\ln^2 N}{N} \right)$ & $\mathcal{O}\left( P_T \frac{\ln^4 N}{N^2} \right)$ \\
			\hline 
			End-Fed PASS & 1 & $\mathcal{O}\left( \frac{\ln^2 N}{N} \right)$ & 0 \\
			\hline 
		\end{tabular}
\end{table}

While the waveguide-based transmission in PASS ensures low-path-loss and reliable signal propagation, it raises a fundamental concern regarding spatial multiplexing gain. This limitation comes from the conventional end-fed PASS architecture, where the waveguide is fed by a single source and the radiated signals are phase-shifted replicas of the same signal. Therefore, a single waveguide in conventional PASS restricts the communication system to $\text{DoF}=1$ and prevents independent multi-stream transmission.

One direct strategy is to deploy multiple waveguides \cite{bereyhi2025mimo,zhao2025pinching}, thereby creating spatially distinct channels to convey independent data streams for multi-user access. This concept of spatial expansion is adopted in SWAN architectures under efficient SS, SA, and SM protocols. However, scaling the number of physical waveguides to accommodate the ultra-dense connectivity envisioned for next-generation networks is practically infeasible.

To address this limitation, the authors of \cite{gan2025c,gan2026center} proposed the novel concept of center-fed PASS (C-PASS) to overcome the $\text{DoF}=1$ limitation of conventional PASS. As illustrated in Fig.~\ref{fig:C-PASS}(a), the wireless signal fed through the input port is bifurcated into two distinct propagation directions along the dielectric waveguide, denoted as forward-propagation (FP) and backward-propagation (BP). According to the performance analysis in \cite{gan2025c}, C-PASS can effectively generate two independent data streams, thereby doubling the available DoFs compared with conventional PASS architectures. 
With a distributed center-feeding architecture~\cite{gan2026center}, the DoF can be further extended to $M$, where $M$ is the number of input ports. As summarized in Table~\ref{tab:analysis_CPASS}, C-PASS can realize a more flexible system by exploiting the additional multiplexing gain, where $N$ is the number of PAs.

To fully exploit the potential of C-PASS, three practical protocols have been proposed in \cite{gan2026center}, namely power splitting (PS), direction switching (DS), and time switching (TS), as illustrated in Fig.~\ref{fig:C-PASS}(b)-(d).

\begin{itemize}
	\item \textbf{Power Splitting:} As illustrated in Fig.~\ref{fig:C-PASS}(b), the PS protocol enables simultaneous signal transmission in both the FP and BP directions. Specifically, the input signal at the $m$-th feed port is bifurcated based on the power splitting ratios $\beta_m^{\text{F}}$ and $\beta_m^{\text{B}}$, which satisfy the power conservation constraint $\beta_m^{\text{F}} + \beta_m^{\text{B}} = 1$, with $\beta_m^{\text{F}}, \beta_m^{\text{B}} \in [0,1]$, $\forall m \in \{1,\ldots,M\}$ and $M$ being the number of feed ports. By allowing the independent optimization of the power splitting ratios for each feed port, PS offers the highest DoF for design and bidirectional coverage. However, this flexible control necessitates precise tuning capabilities at every power splitter, thereby incurring higher hardware implementation complexity.
	
	\item \textbf{Direction Switching:} In contrast to the continuous power division in the PS, the DS protocol partitions the feed ports into two subsets using switches, as depicted in Fig.~\ref{fig:C-PASS}~(c). Specifically, $M_{\text{F}}$ ports are dedicated to exciting the FP direction, while the remaining $M_{\text{B}}$ ports serve the BP direction, satisfying the port conservation constraint $M^{\text{F}} + M^{\text{B}} = M$. Mathematically, this enforces a binary constraint on the splitting coefficients, i.e., $\beta_m^{\text{F}}, \beta_m^{\text{B}} \in \{0,1\}$, subject to $\beta_m^{\text{F}} + \beta_m^{\text{B}} = 1$. Consequently, DS can be theoretically regarded as a special case of PS where the continuous coefficients are discretized into binary ones. Although this reduction in DoF may lead to suboptimal communication performance compared to PS, the DS protocol remains highly attractive for practical deployment. Its reliance on simple ``on-off'' switching mechanisms significantly lowers hardware complexity and cost compared to the precision power splitters required for PS.
	
	\item \textbf{Time Switching:} Diverging from the simultaneous transmission strategies of PS and DS, the TS protocol operates by time-division multiplexing the waveguide resources between the FP and BP directions, as illustrated in Fig.~\ref{fig:C-PASS}~(d). Specifically, all feed ports are switched to serve a single direction at any given instant. Let $\mu^{\text{F}}, \mu^{\text{B}} \in [0,1]$ denote the fractional time allocation ratios for the FP and BP periods, respectively, subject to the constraint $\mu^{\text{F}} + \mu^{\text{B}} = 1$. During each active period, the full power of the feed ports is dedicated to the target direction, meaning the design of the in-waveguide coefficients for FP and BP is effectively decoupled. While this orthogonality reduces the computational complexity of beamforming designs, it imposes stringent requirements for time synchronization.
\end{itemize}

\subsection{Multi-Mode PASS (M-PASS)}
Existing PASS typically relies on dielectric waveguides that support only a single \textit{guided mode}. 
A guided mode refers to an EM field pattern that can propagate along the waveguide with a specific transverse field distribution, corresponding to a physically valid eigenmode solution of Maxwell's equations under the given waveguide geometry, material composition, and boundary conditions.  
Current single-mode PASS typically gives rise to a rank-one in-waveguide channel, which makes it difficult to support multi-user multiplexing over a single waveguide, and thereby fundamentally limits the spatial DoFs and the efficiency of RF resource utilization. 
To overcome this limitation, we introduce the M-PASS \cite{Xu2026MultiModePASS,Xu2026ModeCombiningPASS} in this subsection, which exploits the inherent multi-mode ability within a dielectric waveguide, thus enabling \textit{mode-domain multiplexing} for multi-user communications.

\begin{figure}[t!]
	\centering
	\subfigure[Mode-selective structure.]{
		\includegraphics[width=0.9\linewidth]{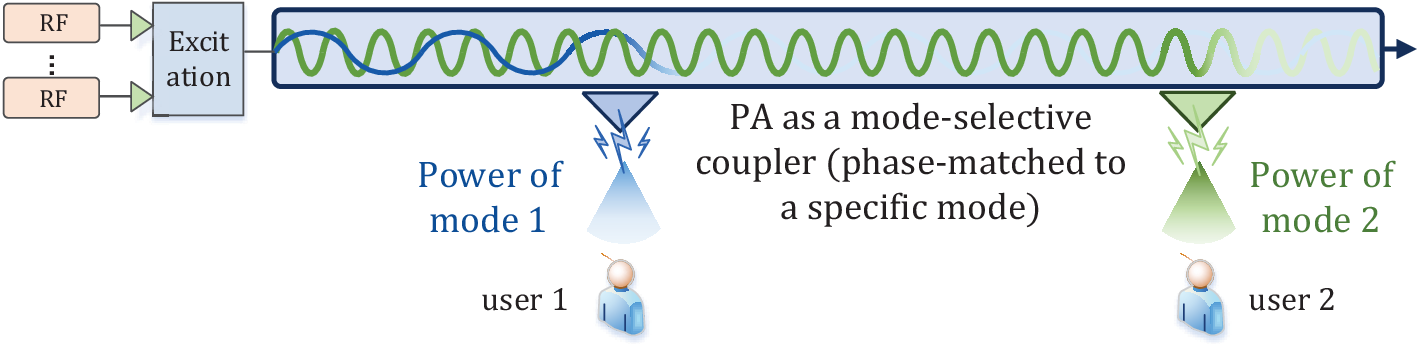}
		\label{fig:multimodePASS1}
	}
	\\
	\subfigure[Mode-combining structure.]{
		\includegraphics[width=0.9\linewidth]{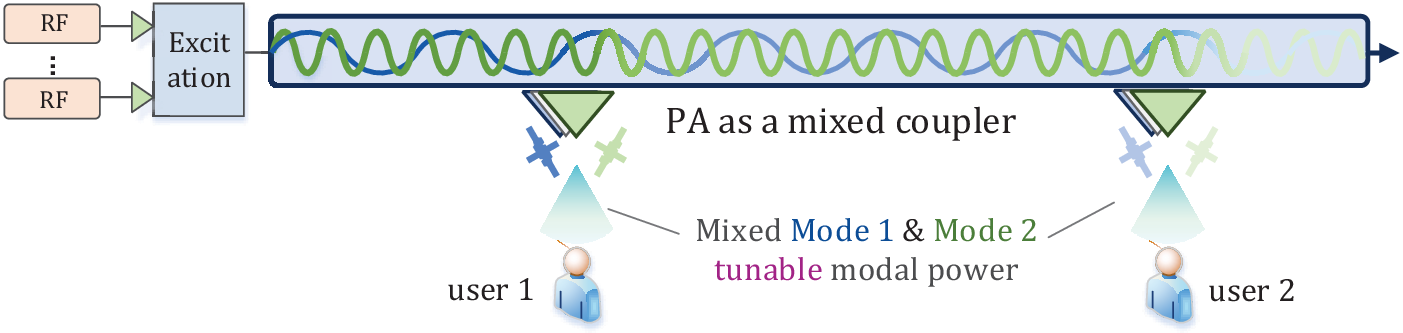}
		\label{fig:multimodePASS2}
	}
	\caption{Two structures for multi-mode PASS implementation.}
    \label{fig:multimodePASS}
\end{figure}

The M-PASS enables the PAs to radiate signals conveyed by multiple guided modes. 
As shown in Fig. \ref{fig:multimodePASS},  the data streams of multiple users are first digitally precoded and launched into the waveguide.  
The waveguide excites $M$ guided modes through multiple feed points or a multi-port launcher. 
These guided modes have different transverse electric field distributions, which can carry independent data streams simultaneously. 
More specifically, each guided mode $m$ is characterized by a distinct propagation constant $\beta_m$. 
For a lossless waveguide, the in-waveguide propagation to PA $n$ located at $x_n$ introduces a mode-dependent phase shift $e^{-j\beta_m x_n}$. 
For a lossy waveguide, this propagation factor can be generalized as $e^{-(\alpha_m+j\beta_m)x_n}$, where $\alpha_m$ denotes the attenuation constant of mode $m$.
Leveraging modal diversity, M-PASS can enhance per-waveguide spatial DoFs.  
The extra spatial DoFs construct a full-rank effective mode-domain channel over a single waveguide, enabling spectral-efficient multi-user communications at a low hardware cost.

To characterize the power radiation of multiple guided modes at the PAs, 
a multi-mode EM coupling model based on CMT has been developed in \cite{Xu2026MultiModePASS}. 
For ease of implementation, each PA is assumed to support a single eigenmode. 
The coupling between the $M$ guided modes and the single-mode PA can then be represented by a system of $M+1$ coupled-mode equations \cite{Xu2026MultiModePASS}. 
When the guided modes are sufficiently separated and the PA-induced inter-mode transfer is weak, this coupled-mode system can be approximately diagonalized into independent two-mode coupling subsystems. 
Accordingly, a closed-form solution can be obtained by extending the conventional two-mode phase-mismatch coupler solution. 
In particular, the radiation power coefficient of guided mode $m$ at PA $n$ generally decreases with their phase mismatch $\Delta\beta_{n,m}$, defined as
\begin{equation}
\Delta\beta_{n,m}=\left|\beta_n^{\mathrm{PA}}-\beta_m\right|,
\end{equation}
where $\beta_{n}^{\mathrm{PA}}$ and $\beta_m$ denote the propagation constants of PA $n$ and guided mode $m$, respectively.

\begin{table*}[!ht]
\centering
\caption{Comparisons of operating protocols for multi-mode PASS.}\label{table:multimodePASS}
\resizebox{\textwidth}{!}{
\begin{tabular}{|l|c|c|c|}
\hline
\textbf{Features}
& \textbf{Mode Selection} 
& \textbf{Mode Combining}
& \textbf{Uniform Mode Combining}\\\hline
PA propagation constant
& Discrete selection: $\beta_n^{\mathrm{PA}}\in\{\beta_1,\ldots,\beta_M\}$
& Continuous tuning: $\beta_n^{\mathrm{PA}}\in[\beta_{\min},\beta_{\max}]$
& Fixed or pre-configured $\beta_n^{\mathrm{PA}}$ \\\hline
EM coupling behavior
& Each PA predominantly radiates one selected mode
& Each PA can radiate multiple modes with adjustable ratios
& Each PA radiates multiple modes with fixed ratios \\\hline
Main design variables
& PA positions, beamforming, and mode assignment
& PA positions, beamforming, and PA propagation constants
& PA positions and beamforming \\\hline
Hardware complexity
& Low-to-moderate, discrete switching among predesigned PA states
& High, continuous or high-resolution tuning required
& Low, pre-configured PA propagation constant \\\hline
Operation flexibility
& Moderate, robust mode isolation but limited modal mixing
& High, flexible multi-mode radiation and interference control
& Moderate, practical modal mixing with a low cost \\\hline
Advantage
& Low complexity and strong mode selectivity
& Highest flexibility and spectral efficiency
& Low-complexity practical implementation \\\hline
Limitation
& Possible leakage and less flexible mode utilization
& Higher calibration and control overhead
& Lower adaptability\\
\hline
\end{tabular}}
\end{table*}

\begin{figure}[h!]
    \centering
    \includegraphics[width=0.8\linewidth]{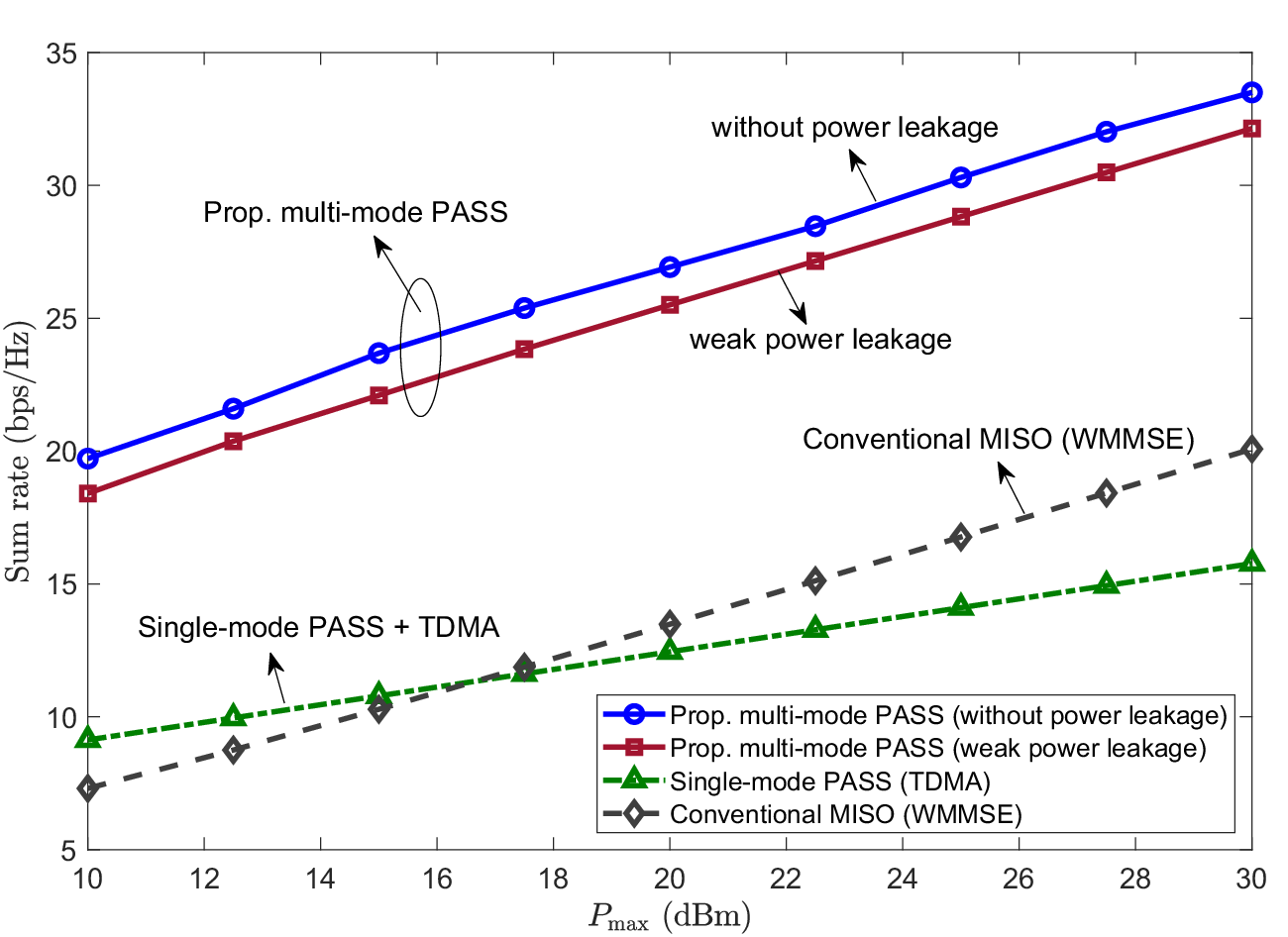}
    \caption{Performance comparisons for mode-selective PASS.}\label{fig:simu_multimodePASS}
\end{figure}

Leveraging the above radiation property governed by phase mismatch, two representative operating protocols have been developed for M-PASS, termed \textit{mode selection} and \textit{mode combining} \cite{Xu2026ModeCombiningPASS}. These two protocols differ fundamentally in how the PA propagation constants are configured and, consequently, how each PA couples to multiple guided modes.
\begin{itemize}
    \item \textbf{Mode selection}: 
    In the mode selection protocol, each PA is configured to predominantly radiate the signal of one selected guided mode. 
    This is realized by choosing the PA propagation constant from a discrete set: 
    \begin{equation}
    \beta_n^{\mathrm{PA}}\in \{\beta_1,\beta_2,\ldots,\beta_M\}.
    \end{equation}
    If PA $n$ selects mode $m$, then $\beta_n^{\mathrm{PA}}=\beta_m$ and $\Delta\beta_{n,m}=0$, which maximizes the radiation power of the selected mode. 
    Meanwhile, the radiation power of the other non-selected modes is suppressed due to phase mismatch. 
    
    A practical implementation of mode selection is PA grouping, where PAs are divided into multiple groups, and each group is phase-matched to a dedicated guided mode. 
    Depending on whether the PAs leak power of the unmatched guided modes, the resulting system operates under either a non-leakage regime or a weak-leakage regime. 
    Fig. \ref{fig:simu_multimodePASS} provides numerical results for a two-user two-mode PASS equipped with $12$ PAs over a $20$-meter circular waveguide with a radius of 5 mm. The system operates at $28$ GHz, and the waveguide supports quasi-TE$_{11}$ and quasi-TM$_{01}$ modes with propagation constants $\beta_{1}=456.399$ m$^{-1}$ and $\beta_{2}=335.527$ m$^{-1}$. The PAs are divided into two groups, each alternately matched to one of the supported modes. For both non-leakage and weak-leakage regimes, M-PASS achieves a higher spectral efficiency compared to conventional single-mode PASS using time-division multiple access (TDMA) and hybrid beamforming-based multiple-input single-output (MISO) system. This demonstrates the effectiveness of mode-domain spatial multiplexing for M-PASS. 
    \item \textbf{Mode combining}: 
    In contrast to mode selection, mode combining allows one PA to extract and radiate signal power from multiple guided modes by tuning its propagation constant, without necessarily matching one specific mode. 
    Specifically, the PA propagation constant is continuously tunable within a feasible range:
    \begin{equation}
    \beta_n^{\mathrm{PA}}\in[\beta_{\min},\beta_{\max}].
    \end{equation}
    By properly configuring $\beta_n^{\mathrm{PA}}$, the phase mismatches $\Delta\beta_{n,m}$ can be adjusted, thereby controlling the radiation power coefficients of multiple modes at PA $n$. 
    As a result, each PA acts as a programmable multi-mode coupler and provides a higher degree of flexibility for multi-user interference management and beam shaping. 
    However, the flexibility of mode combining leads to a more challenging joint optimization problem, where the transmit beamforming, PA positions, and PA propagation constants need to be optimized simultaneously. 
    
    A low-complexity implementation is \textit{uniform mode combining}. 
    For a two-mode PASS, all PAs can adopt the same pre-configured propagation constant: 
    \begin{equation}
    \beta_1^{\mathrm{PA}}
    =
    \beta_2^{\mathrm{PA}}
    =
    \cdots
    =
    \beta_N^{\mathrm{PA}}
    =
    \frac{\beta_1+\beta_2}{2}.
    \end{equation}
    This design avoids real-time manipulation of $\{\beta_n^{\mathrm{PA}}\}$ and significantly reduces the hardware complexity, while still enabling each PA to radiate a mixture of multiple guided modes. 
    For more than two modes, a group-wise uniform mode-combining design can be adopted, where different PA groups are pre-configured to combine adjacent modes.
\end{itemize}

Table \ref{table:multimodePASS} summarizes the key features of the above two structures. Note that these two structures can be applied in different scenarios.  
Specifically, the \emph{mode-selective} structure is most effective when the guided modes exhibit strong inherent separability and when the deployment environment is relatively static (e.g., corridors) and users are sparsely distributed. In these cases, the selective phase matching at each PA offers robust and low-complexity  spatial multiplexing. In comparison, the \emph{mode-combining} structure is preferred when high coupling flexibility and real-time adaptation are required. The tunability of the modal signal power at each PA makes it suitable for dense or rapidly changing environments (e.g., urban hot spots, IoT) or when the guided modes are closely spaced or near-degenerate. In such scenarios, dynamic modal configuration and adaptive beam shaping become essential to improve multi-user system performance.

\subsection{Discussion and Outlook}
To fully utilize the flexibility of PASS, it is necessary to develop new models and designs to support the implementation of these PASS variants.
\subsubsection{New Modeling and Optimization Frameworks for PASS Variants}
This section discusses several new variants of PASS, outlining their underlying rationales and associated advantages.
However, characterizing the potential of these PASS variants requires developing new physics-based models to accurately represent in-waveguide channels.
This step is vital for further optimization and analysis.
Furthermore, realizing the full potential of PASS necessitates novel optimization frameworks tailored to these emerging architectures.

\subsubsection{Wireless Feeding for PASS}
While PASS enables signal delivery via LoS propagation, its deployment is limited by environmental factors, such as building obstructions and terrain irregularities.
In such scenarios, replacing the wired connection between the base station and waveguide feed points with wireless links can be a favorable choice.
This architecture enhances the flexibility and adaptability of PASS deployments \cite{wijewardhana2025wireless}.
To capitalize on these advantages, future research should prioritize protocol design and explore both centralized and distributed approaches for wireless feeding control.

\section{PASS-Assisted Sensing} \label{sect:communication_to_sensing}
Due to channel reconfigurability, the communication benefits of PASS are evident and have been extensively studied in prior work. 
However, as another key dimension, the sensing capability of PASS remains underexplored. 
In particular, the flexibility of PAs can enhance the sensing performance, especially given the double-fading effect associated with echo-signal reception.
In what follows, we first explain the motivation behind adopting PASS for sensing.
Then, we present several designs for PASS-assisted sensing that encompass sensing, localization, and user tracking.
Finally, we discuss the interplay between sensing and communications, which is required for integrated sensing and communications (ISAC) networks.

\subsection{Motivation of PASS-Assisted Sensing}
In fact, the benefits of PASS-assisted sensing are threefold: \emph{Near-Field Benefits}, \emph{Enhanced Sensing Resolution}, and \emph{Line-of-Sight Creation}. 
First, due to its large aperture, PASS can leverage the near-field phenomenon, enabling comprehensive acquisition of sensing parameters.
In particular, the boundary of the near-field region is defined by the Rayleigh distance, which can be mathematically expressed as $2D^2/\lambda$ with $D$ denoting the aperture size of antenna arrays and $\lambda$ being the wavelength \cite{liu2025near}.
Unlike ELAAs, which typically have apertures of several meters, PASS can extend the aperture to tens of meters.
As such, the spherical wavefronts have a non-negligible effect on the impinging sensing signal, i.e., uplink pilots or echo signals.
Leveraging the curvature of spherical waves, the polar-domain coordinates and velocities of the sensing targets can be readily obtained \cite{jiang2025near}. 
More importantly, unlike conventional ELAAs, which create a near-field region spanning tens of meters with hundreds of antenna elements, PASS creates a vast near-field region with only a handful of PAs, thereby reducing signal processing complexity. 
The second advantage of PASS’s large aperture is enhanced resolution. 
More specifically, the sensing resolution is directly determined by the aperture size rather than the number of antennas \cite{wang2025near}. 
As such, PASS can significantly enhance sensing resolution and is a promising candidate for sensing/localization tasks \cite{wang2025sensing}.
Third, the LoS propagation conditions created by PASS are favorable for sensing, since LoS links have high SNR and can directly relate sensing parameters, such as positions and velocities, to the received signal without complex models of reflections and deflections from NLoS links.
Moreover, as next-generation telecommunication networks are anticipated to operate in high-frequency bands, NLoS links are not sufficiently strong or robust for sensing tasks due to atmospheric-induced attenuation \cite{wang2018millimeter}.
In this case, thanks to its ability to create LoS paths, PASS can provide consistent, robust sensing channels.

\subsection{Design of PASS-Assisted Sensing}
To harness the theoretical benefits of PASS-assisted sensing, we discuss representative application scenarios to highlight its advantages, including pure sensing, localization, and user tracking. 

\subsubsection{Sensing}
Given the pure sensing functionality, our objective is to answer the following question: What are the theoretical sensing performance limits? 
Existing studies addressing this question can be divided into two categories: Sensing SNR and Cramér–Rao lower bound (CRLB). 
Specifically, sensing SNR reflects the received/reflected power associated with the target \cite{zhang2025integrated}. 
The authors of \cite{zhang2025integrated, qin2025joint, hao2025segmented} showed that large-scale fading reconfigurability can mitigate the double-fading effects of round-trip sensing channels and theoretically increase sensing SNR. 
In addition to SNR, the CRLB can serve as a sensing performance metric that can be physically interpreted as a fundamental lower bound on the error variance incurred by any unbiased estimator \cite{liu2022integrated,jiang2025cramer}. 
Thus, the CRLB is more directly related to sensing performance. 
In other words, minimizing the CRLB equivalently reduces the estimation error variance and improves sensing accuracy. 
To this end, the authors of \cite{ding2025pinching} derived the CRLB for an uplink PASS-based sensing setup. 
Another study addressed CRLB minimization in \cite{wang2025sensing} using a particle swarm optimization (PSO)-based algorithm. 
This further demonstrates that PASS-assisted sensing is more robust to the errors caused by initial sensing parameter estimation. 
To reduce the dependence on the true sensing parameter values in the expressions for the CRLB, the authors of \cite{jiang2025pinching_sensing} analyzed uplink PASS sensing from a Bayesian CRLB perspective, leveraging prior knowledge of target position distributions.

\subsubsection{Localization}
PASS enables high-precision user localization by activating multiple PA elements or subarrays that are spatially distributed with respect to the targets. The authors of \cite{Zhang2025PASSIndoor} first explored PASS-based indoor positioning using received signal strength indicator (RSSI) measurements. Since the PASS channel response is inherently dependent on the PA positions, it can be modeled as a superposition of spherical-wave components associated with the line-of-sight (LoS) path and multiple non-line-of-sight (NLoS) scatterers \cite{Xu2025PASSLocalization, zhou2025channel}. Motivated by this property, the authors of \cite{Xu2025PASSLocalization} investigated the joint estimation of user and scatterer locations for spherical-wave channel reconstruction in PASS. Both single-waveguide and multi-waveguide architectures were considered. In the single-waveguide structure, multiple colinear subarrays are activated along a single waveguide in a time-division manner, whereas the multi-waveguide structure allows multiple subarrays on different waveguides to be activated simultaneously for channel probing. A geometry-consistent orthogonal matching pursuit (GCL-OMP) algorithm was proposed that exploits OMP for local direction estimation at each subarray and incorporates multi-subarray geometric relationships to achieve centimeter- and decimeter-level localization accuracy in 2D and 3D scenarios. 
Furthermore, the CRLB analysis reveals that localization accuracy improves with increasing geometric diversity, i.e., the spatial spread of observation directions across multiple subarrays.

\subsubsection{User Tracking}
Another application scenario for PASS is user tracking. 
Unlike localization tasks, where sensing targets are static or low-mobility, user tracking aims to track moving users by exploiting Doppler frequencies \cite{liu2020radar}.
Beyond conventional far-field sensing scenarios, PASS, endowed with near-field benefits, can extract the full-dimensional mobility status of sensing targets, including polar-domain positions and velocities \cite{jiang2025near}. 
As such, the complete target trajectory can be captured simply using the near-field channel feature, thereby eliminating the need for prior knowledge of the target's trajectory. 
To harness these benefits, the authors of \cite{jiang2025pinching} proposed a PASS-based sensing-aided covert communication scheme that leverages an extended Kalman filter (EKF)-based warden-tracking method to consistently acquire the malicious user's channel state information.
More importantly, PASS can create a vast near-field region with a handful of PAs.
Hence, the near-field phenomenon is enhanced without requiring hundreds or even thousands of antenna elements.

\subsection{Interplay Between Sensing and Communications of PASS}
Since the superiority of PASS-assisted sensing has been confirmed by the above works, it is also important to investigate the interplay between communication and sensing functionalities, which is highlighted in IMT2030 \cite{kaushik2024toward}.
There are two types of ISAC systems: Sensing-aided communication and communication for sensing.

\begin{figure}[t!]
    \centering
    \includegraphics[width=0.8\linewidth]{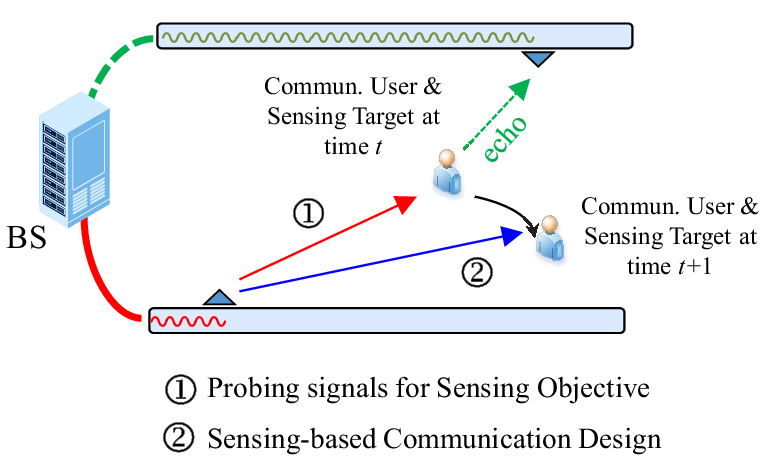}
    \caption{Illustration of sensing-aided communications.}
    \label{fig:sensing_for_communications}
\end{figure}
\subsubsection{Sensing-Aided Communications}
In this category, the sensing functionality is used as a perception tool to gather environmental information.
Then, these sensing results will be exploited to support the design of wireless communication systems, making communication technologies less computationally intensive and more responsive.
This paradigm is illustrated in Fig. \ref{fig:sensing_for_communications}, where sensing results in the $t$-th time slot facilitate the transmission design for the upcoming $(t+1)$-th time slot.
In particular, the authors of \cite{gan2025llm} developed a large-language model (LLM)-based learning framework to perform beam training.
In this method, multimodal sensing information, including camera images, light detection and ranging (LiDAR), and global positioning system (GPS) data, is used for beam training, serving as an alternative approach for CSI acquisition. 
An LLM-based supervised learning mechanism is devised to jointly perform codebook generation and beam selection.
Moreover, the authors of \cite{zhou2025channel} devised a sensing method to obtain the environmental information in a PASS-assisted communication network.
Building on this sensing information, the CSI can be recovered to achieve continuous channel estimation for PASS.

\subsubsection{Communication for Sensing}
\begin{figure}[t!]
    \centering
    \includegraphics[width=0.8\linewidth]{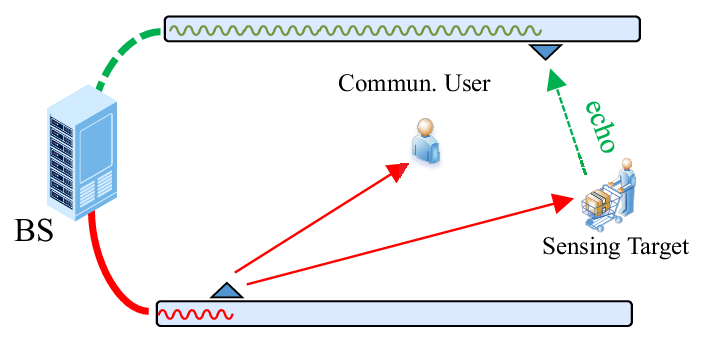}
    \caption{Illustration of communication for sensing.}
    \label{fig:communication_for_sensing}
\end{figure}
The other category of ISAC systems is communication for sensing, which leverages communication signals for fulfilling sensing requirements.
As such, the sharing between radio resources and hardware improves the spectrum- and hardware efficiency \cite{liu2020joint, liu2022survey_isac}.
This paradigm is illustrated in Fig. \ref{fig:communication_for_sensing}, where the communication signals are leveraged for sensing the target.
Within this category, the authors of \cite{qin2025joint} and \cite{zhang2025integrated} were the first to investigate a PASS-assisted ISAC system, where the communication objective is met while maintaining the sensing SNR above a specified threshold.
Additionally, the authors of \cite{bozanis2025cramer}, \cite{li2025pinching}, and \cite{khalili2025pinching} incorporated sensing accuracy directly into ISAC designs by accounting for the round-trip sensing channel and employing uniform linear arrays (ULAs) for echo reception.
To mitigate signal re-radiation issues in the PASS uplink, the authors of \cite{hao2025segmented, jiang2026swan_isac_conf} investigated segmented waveguide-based ISAC systems.
Furthermore, the segmented structure mitigates in-waveguide propagation losses by reducing the propagation distance within each waveguide segment.
This approach theoretically demonstrates the advantages of SWAN-assisted sensing and identifies the trade-off between sensing and communication through optimization.
From an information-theoretic perspective, the authors of \cite{ouyang2025rate} analyzed the rate region of PASS-aided ISAC systems, revealing a fundamental trade-off between communication rate and sensing rate. 

\subsection{Discussion and Outlook}
Despite the fruitful research on PASS for sensing, many open questions and potential extensions remain, and we discuss some of them below.

\subsubsection{Sensing with New Types of PASS}
Currently, most existing work is based on the conventional PASS architectures, in which received signals are automatically aggregated within the waveguide.
Consequently, as the signal propagates to the feed point, it will reradiate from the PAs, making uplink signal modeling intractable. 
Therefore, adopting new types of PASS, such as SWANs, can simplify the uplink signal modeling.
Moreover, the compatibility of sensing and communication across a wide range of PASS architectures also warrants future research.  

\subsubsection{Machine Learning for Sensing}
Exploring machine learning for PASS-assisted sensing can also be an important research direction.
Due to the presence of echo signals in sensing, the signal model is complex, leading to a highly multimodal optimization space.
This issue makes the use of convex optimization tools challenging.
Additionally, existing work mostly relies on an element-wise search method to optimize PA positions, resulting in high computational complexity.
Machine learning can potentially be used to reduce the high complexity of PASS-aided sensing.

\section{Performance Analysis of PASS} \label{sect:preformance_analysis}
In this section, we review the research progress on the communication and sensing performance of PASS.
The performance advantage of PASS over conventional fixed-antenna systems arises from the extended coverage provided by the dielectric waveguide and the flexibility in positioning the PAs. This advantage becomes particularly pronounced when the randomness of user locations is taken into account.
For communication performance, we focus on two key metrics: the achievable user rate and the outage probability (OP). 
For sensing performance, we examine studies that analyze the CRLB and sensing SNR.

\subsection{Communication Performance}
This subsection reviews existing studies that analyze the communication performance of PASS. An important research direction is to apply stochastic geometry to evaluate the average network performance under random user distributions. 
Against this background, two performance metrics have attracted particular attention: the \emph{average achievable rate} and the \emph{OP}.
\subsubsection{Achievable Rate}
The authors in \cite{ding2025flexible} analyzed the average user rate achieved by PASS in a system where a single waveguide is placed at the center of a square service region, and the user is uniformly distributed within that region. To maximize the downlink received SNR, the PA is positioned at the projection of the user's location onto the waveguide. This study provides closed-form expressions for the average user rate and its high-SNR approximation, and shows that PASS outperforms conventional fixed-antenna systems in terms of average user rate.

The analysis in \cite{ding2025flexible} is based on an idealized model that ignores waveguide hardware impairments by neglecting in-waveguide propagation loss. As an extension, the authors of \cite{xu2025pinching1} determined the optimal PA position while incorporating in-waveguide propagation loss under the same system configuration. They further derived closed-form expressions for the average user rate and showed that in-waveguide propagation loss has a negligible effect on PASS performance, implying that it can be safely omitted for system-level analysis.

The studies discussed above determine the optimal PA location under the assumption that the PA can be activated at any point along the waveguide. From a practical deployment perspective, however, it is often more realistic to allow PA activation only at a set of predefined positions, which reduces the deployment and activation cost of PASS. Considering this constraint, the authors of \cite{tyrovolas2025ergodic} analyzed the average user rate of a single-user single-PA PASS setup in which the PA can only be activated at several uniformly spaced predefined locations along the waveguide. Beyond the common center-deployed waveguide configuration, the authors of \cite{zhang2025performance} examined the average rate when the waveguide is placed along the edge of the region or along its diagonal, showing that waveguide placement has a substantial impact on achievable-rate performance.

All of the works discussed above study the use of a single waveguide to serve one user. To further exploit the array gain enabled by multiple PAs, the study in \cite{ouyang2025array} examined a PASS configuration in which several PAs are deployed along a single waveguide to serve a single user. The locations of the PAs are optimized using the position-refinement method introduced in \cite{xu2025pinching} to maximize the downlink received SNR. Based on this optimization, the authors derived closed-form expressions for the SNR and demonstrated that the user rate achieved by PASS does not necessarily increase monotonically with the number of activated PAs.

However, this study does not account for user mobility or spatial randomness. Building on \cite{ouyang2025array}, the authors of \cite{hou2025performance} incorporated randomness in the user's location and derived closed-form expressions for the average user rate. A further extension to a scenario in which multiple waveguides, each with multiple PAs, serve a single user is presented in \cite{cheng2025performance}. This study analyzed the achievable rate when several waveguides are connected to a phase shifter network and multiple RF chains to enable tri-hybrid beamforming. Additionally, the study in \cite{zhou2025joint} investigated a system in which multiple waveguides and a BS serve a single user, and analyzed the average user rate under different cooperation strategies between the BS and the waveguide-mounted PAs.

Extensions of achievable-rate analysis to multiuser PASS have also been investigated. The study in \cite{ouyang2025capacity} characterized the achievable rate region for two-user downlink and uplink PASS, where multiple PAs on a single waveguide serve both users. The work in \cite{biting2025noma} examined the achievable rate when a single waveguide is combined with partial NOMA to serve two users by dynamically controlling the signal overlap ratio between the near and far users. Building on this line of research, the study in \cite{qian2025pinching} extended the analysis to a general multiuser setting by evaluating the achievable rate of an MRT beamforming-assisted multi-waveguide PASS, in which each waveguide employs one PA positioned near its associated user.

The work in \cite{ding2025los} further considered both the mitigation of large-scale path loss through PA deployment and the impact of LoS blockage. Specifically, the PA positions are optimized along the waveguides so that desired links maintain strong LoS paths while environmental blockages, such as walls or obstacles, are deliberately exploited to attenuate interference links. This approach provides new insights into PA placement for exploiting environmental features to enhance multiuser access, a concept related to EDMA \cite{wang2025pinching,ding2026environment}. The multiuser PASS studies discussed above primarily focus on unicast transmission. In contrast, the work in \cite{shan2025exploiting} analyzed the average user rate achieved by a single waveguide serving two multicast users.

A summary of the main contributions to achievable-rate analysis for PASS is given in Table \ref{Table: PASS_Rate_Analysis}.

\begin{table*}[!t]
\centering
\caption{Contributions to achievable-rate analysis of PASS}
\resizebox{0.95\textwidth}{!}{
\begin{tabular}{|l|l|l|l|l|l|}
\hline
\textbf{Ref.} & \textbf{Waveguides}                                     & \textbf{PAs per Waveguide} & \textbf{Users}     & \textbf{Direction} & \textbf{Characteristics} \\ \hline
\cite{ding2025flexible}   & Single     & Single                & Single   & DL  & Closed-form average rate with high-SNR approximation for ideal waveguide               \\ \hline
\cite{xu2025pinching1}    & Single     & Single                & Single   & DL  & Optimal PA placement with in-waveguide loss showing negligible performance impact               \\ \hline
\cite{tyrovolas2025ergodic}   & Single     & Single                & Single   & DL  & Average rate with PA activation restricted to predefined discrete locations              \\ \hline
\cite{zhang2025performance}   & Single     & Single                & Single   & DL  & Average rate under various waveguide placements within the service region              \\ \hline
\cite{ouyang2025array}   & Single     & Multiple                & Single   & DL  & SNR-based PA-location optimization revealing non-monotonic rate scaling              \\ \hline
\cite{hou2025performance}   & Single     & Multiple                & Single   & UL  & Average rate with random user locations for multi-PA PASS             \\ \hline
\cite{cheng2025performance}   & Multiple     & Multiple                & Single   & DL  & Achievable rate of tri-hybrid beamforming using multi-waveguide deployment            \\ \hline
\cite{zhou2025joint}   & Multiple     & Multiple                & Single   & DL  & Average rate for joint BS-waveguide transmission with cooperation strategies           \\ \hline
\cite{ouyang2025capacity}   & Single     & Multiple                & Two   & DL/UL  & Achievable-rate region for two-user PASS with multi-PA waveguide           \\ \hline
\cite{biting2025noma}   & Single     & Single                & Two   & DL  & Achievable rate for partial NOMA with dynamic signal-overlap control           \\ \hline
\cite{qian2025pinching}   & Multiple     & Single                & Multiple   & DL  & Multiuser rate of MRT-assisted PASS with per-waveguide nearest-user assignment           \\ \hline
\cite{ding2025los}   & Multiple     & Single                & Multiple   & DL  & ZF-based multiuser rate with blockage-assisted interference suppression via EDMA           \\ \hline
\cite{shan2025exploiting}   & Single     & Single                & Two (multicast)   & DL  & Multicast achievable rate for two users served by a single waveguide           \\ \hline
\end{tabular}}
\label{Table: PASS_Rate_Analysis}
\end{table*}

\subsubsection{Outage Probability}
Another important metric for evaluating network performance is the \emph{OP}, which is defined as the likelihood that a user fails to achieve a target SNR. Its complement, obtained by subtracting the OP from one, represents the probability that the user meets the target SNR, which we refer to as \emph{coverage probability}. The authors in \cite{tyrovolas2025performance} derived closed-form expressions for the OP of a single-user PASS employing a single waveguide. 
Their analysis considered both an ideal lossless waveguide and a practical lossy waveguide. This work was later extended to a wireless-powered PASS in \cite{cao2025performance}, where the user first harvests energy from the downlink signal transmitted by one energy-providing waveguide and then communicates with another uplink waveguide. The study in \cite{zhang2025performance1} further extended \cite{tyrovolas2025performance} by deriving closed-form OP expressions for a single-user single-PA PASS under both full-coverage and partial-coverage waveguide configurations, with and without propagation loss. These single-user analyses were generalized to a two-user case in \cite{cheng2025ontheperformance}, where both OMA and power-domain NOMA were considered to mitigate inter-user interference. A subsequent extension in \cite{tegos2025uplinkrsma} analyzed a two-user, two-PA uplink PASS using rate-splitting multiple access (RSMA). Closed-form OP expressions were derived to show that RSMA provides a more robust interference-management framework than NOMA.

A more general scenario involving multiple waveguides and multiple users was examined in \cite{wang2025pinchingImmersive}, where waveguide division multiple access (WDMA) was introduced. In this setup, each waveguide serves one user, and an integral-form expression for the OP was derived by incorporating the randomness of user locations. Extensions of the OP analysis to multicell PASS setups were presented in \cite{sun2025stochastic} and \cite{zhu2025topological}. These studies used \emph{stochastic-geometry tools} to evaluate the coverage probability in the presence of both inter-cell and intra-cell interference. In particular, the authors of \cite{sun2025stochastic} focused on interference from other waveguides under LoS blockage, whereas those of \cite{zhu2025topological} also considered interference from conventional fixed-antenna BSs.

A summary of the main contributions to OP analysis of PASS is given in Table \ref{Table: PASS_OP_Analysis}.

\begin{table*}[!t]
\centering
\caption{Contributions to OP analysis of PASS}
\resizebox{0.95\textwidth}{!}{
\begin{tabular}{|l|l|l|l|l|l|}
\hline
\textbf{Ref.} & \textbf{Waveguides}                                     & \textbf{PAs per Waveguide} & \textbf{Users}     & \textbf{Direction} & \textbf{Characteristics} \\ \hline
\cite{tyrovolas2025performance}   & Single     & Single                & Single   & DL  & Closed-form OP for ideal and lossy waveguides               \\ \hline
\cite{cao2025performance}    & Single     & Single                & Single   & UL  & OP of wireless-powered PASS with energy harvesting               \\ \hline
\cite{zhang2025performance1}   & Single     & Single                & Single   & DL  & OP for full and partial waveguide coverage with propagation loss              \\ \hline
\cite{zhang2025performance}   & Single     & Single                & Single   & DL  & Average rate under various waveguide placements within the service region              \\ \hline
\cite{cheng2025ontheperformance}   & Single     & Single                & Two   & DL  & Two-user OP under OMA and NOMA transmission             \\ \hline
\cite{tegos2025uplinkrsma}   & Single     & Two                & Two   & UL  & OP of RSMA-based uplink PASS with improved interference robustness             \\ \hline
\cite{wang2025pinchingImmersive}   & Multiple     & Single                & Multiple   & DL  & OP of WDMA with per-waveguide user assignment and random user locations            \\ \hline
\cite{sun2025stochastic}   & Multiple (multicell)     & Single                & Multiple (multicell)   & DL  & OP under multicell interference with LoS blockage using stochastic geometry           \\ \hline
\cite{zhu2025topological}   & Multiple (multicell)     & Single                & Multiple (multicell)   & DL & OP under interference from waveguides and fixed-antenna BSs           \\ \hline
\end{tabular}}
\label{Table: PASS_OP_Analysis}
\end{table*}

Collectively, these works demonstrate the capability of PASS to enhance the user channel gain and reduce inter-user interference. They highlight the potential of PASS to improve both the achievable rate and the outage performance, which ultimately contributes to improved network coverage and throughput.

\begin{table*}[!t]
\centering
\caption{Contributions on sensing performance analysis for PASS. Here, ``UL" stands for uplink sensing, ``DL" stands for downlink sensing, and ``DL\&UL" stands for downlink probing signal transmission and uplink echo reception. }
\resizebox{0.95\textwidth}{!}{
\label{Table:PASS_sensing_analysis}
\begin{tabular}{|l|l|l|l|p{0.7\textwidth}|}
\hline
\textbf{Ref.} & \textbf{Metric} & \textbf{Direction} &\textbf{Array Type} & \textbf{Main Contributions} \\ \hline
\cite{ding2025analytical} & CRLB & UL & PASS & Fundamental sensing-communication trade-off in PASS. \\ \hline
\cite{bozanis2025cramer}  & CRLB & DL\&UL & Tx: PASS, Rx: ULA & Theoretical sensing-accuracy gain (up to $10\times$) compared to conventional FPA. \\ \hline
\cite{jiang2025pinching_sensing} & BCRLB & UL &PASS & The misalignment between the sensing-sensitive centroid and the user-distribution centroid. \\ \hline
\cite{he2025pinching} & CRLB & DL &PASS  &Stochastic-geometry-based analysis of PASS sensing performance. \\ \hline
\cite{ouyang2025rate} & Sensing Rate & DL\&UL & PASS & Derives the Pareto front for the sensing-communication trade-off. \\ \hline
\cite{hao2025segmented} & Sensing SNR & DL\&UL & Segmented PASS & Sensing performance enhancement under segmented structures. \\ \hline
\cite{jiang2026pinching} & ZZB & UL & PASS & Derivation of ZZBs and asymptotic results; Surrogate function for ZZB minimization. \\ \hline
\end{tabular}}
\end{table*}

\subsection{Sensing Performance}
Unlike the analysis of PASS-based communications, sensing performance is still underexplored.
As a performance metric related to sensing accuracy, the CRLB is most frequently used in sensing performance analysis.
As a pioneering work, the authors of \cite{ding2025analytical} derived the expression of the CRLB and unveiled the superiority of PASS-based sensing.
The findings of this study can be summarized in two key aspects: First, PASS can offer a uniform sensing performance for targets that may not be uniformly distributed; and second, directly placing PAs above the sensing target does not yield the lowest CRLB, thus indicating the trade-off between PASS-based communications and sensing.
As an extension, the authors of \cite{bozanis2025cramer} derived the CRLB for a round-trip scenario, where the probing signal is sent by PASS while the echo signal reflected from the target is received by a conventional uniform linear array (ULA).
Their findings showed that, with the same number of PAs, PASS can achieve a tenfold improvement in sensing performance.
In contrast to the conventional CRLB, the Bayesian CRLB (BCRLB) does not require evaluation at unknown sensing-parameter values; instead, BCRLB computes an average over the prior distribution of these parameters.
Motivated by this, the authors of \cite{jiang2025pinching_sensing} used the Bayesian CRLB as a performance metric, leveraging prior knowledge of the distributions of the sensing targets to mitigate dependence on specific locations.
The results of this paper demonstrated the misalignment between the sensing centroid (i.e., the optimal PA position that minimizes BCRLB) and the distribution centroid (i.e., the center of the target's prior distribution), and provided the rationale behind this observation based on the near-field effects caused by PASS.
In a separate study, the authors of \cite{he2025pinching} derived the CRLB distribution using stochastic geometry and elucidated the fundamental limits of PASS-based localization systems.
In addition to CRLB, Ziv-Zakai bound (ZZB) is also exploited, which provides a tight lower bound on the Bayesian MSE for a wide SNR range \cite{zhang2023ziv_zakai}.
Therefore, the authors of \cite{jiang2026pinching} investigated ZZB in PASS-assisted sensing and revealed the issue of sensing ambiguity.
Furthermore, the sensing rate is also a widely used performance metric that quantifies the information-theoretic limit of this sensing task and is defined by the sensing mutual information (MI).
Based on this metric, the authors of \cite{ouyang2025rate} derived closed-form expressions for the achievable communication and sensing rates of a PASS-based ISAC system.
Building on this, the Pareto front was identified, revealing the trade-off between sensing and communication.
Moreover, the authors of \cite{hao2025segmented} defined the sensing performance gain as the ratio of the sensing SNRs of PASS achieved with and without a segmented structure.
Their analytical results demonstrated the benefits of a segmented structure for mitigating in-waveguide loss and revealed how this gain scales up with the number of segments.
A summary of the analysis of the sensing performance of PASS is provided in Table \ref{Table:PASS_sensing_analysis}.

\subsection{Discussion and Outlook}
Based on this section, some open challenges and possible extensions of the existing works are discussed in what follows: 

\subsubsection{Performance Analysis on New Types of PASS} Current research efforts focus on the usage of conventional PASS for probing signal transmission and the echo signal reception.
Recently, new types of PASS, such as SWAN, C-PASS, and M-PASS, have been proposed, which are detailed in Section \ref{sect:promising}. 
Although these PASS variants offer improved performance or reduced implementation costs, their architectures differ from that of the conventional PASS. As a result, the current analytical frameworks cannot adequately characterize their performance.
Future research should address this gap by developing new approaches to more effectively evaluate their potential for wireless sensing.

\subsubsection{Performance Analysis on Sensing and Communication Tradeoff} Previous work has shown that the target's distribution centroid does not overlap with the sensing-sensitive region.
Placing PAs directly above the sensing target does not result in optimal sensing performance; however, this configuration can enhance the communication rate by reducing path loss.
Therefore, studying the trade-off between sensing and communication is an important issue for the performance analysis of PASS-enabled ISAC systems.

\section{Advanced Mathematical Tool for PASS: From Optimization to ML}\label{sect:optimization}
In this section, we provide a comprehensive review of advanced mathematical tools for PASS from two complementary perspectives, namely, mathematical optimization and machine learning. We first identify key challenges in PASS design and present a systematic taxonomy of representative mathematical optimization methods. We then shift our focus to machine learning techniques for PASS and discuss how such techniques can be leveraged to enable real-time control in large-scale PASS deployments.

\subsection{A Mathematical Optimization Perspective}\label{sec:optimization}
In general, the optimization of PASS (e.g., in terms of spectral/energy efficiency, sensing performance, and secure communications) is very challenging for the following reasons:
\begin{itemize}
    \item\textbf{Intrinsic non-convexity:} The locations of PAs jointly shape the signal phase and path loss in PASS, which gives rise to the pinching beamforming design problem. However, variations in PA positions can induce rapid phase changes and significantly affect the path loss of spherical-wave channels, resulting in highly oscillatory and strongly non-convex objective functions. As a result, most PASS optimization problems cannot be handled by standard convex-optimization tools.
    \item\textbf{Strong coupling:} PASS typically relies on the joint optimization of multiple PAs' positions, the transmit/hybrid beamforming, and power allocation among waveguides/PAs. 
    This leads to a highly coupled optimization problem. 
    \item\textbf{High computational complexity and scalability issues:} As the number of PAs, users, and waveguides increases, the joint beamforming optimization for PASS becomes extremely high-dimensional. 
    Exhaustive search is infeasible, and efficient solutions demand careful exploitation and decomposition of the problem structure to ensure scalability.
\end{itemize}

To address these challenges, existing PASS optimization approaches can be broadly divided into three major categories, as summarized in Table \ref{table:optimization_methods} and discussed below.

\subsubsection{Structure-based Optimization Methods}
Structure-based optimization methods explicitly exploit the mathematical structure of PASS optimization problems, such as phase alignments, Karush–Kuhn–Tucker (KKT) optimality conditions, block variable decomposition, or surrogate upper/lower bounds, to derive deterministic iterative updates with provable convergence guarantees.

\begin{itemize}
    \item \textbf{Globally optimal algorithms:} For a single-waveguide PASS, the authors of \cite{ding2025analytical} analytically characterized the optimal PA placement for PASS under both OMA and NOMA transmission strategies. 
    For a general multi-waveguide PASS, the joint optimization problem of transmit beamforming and PA locations is highly non-convex and has multiple local optima \cite{xu2025pinching,Xu2026ModeCombiningPASS}, because the phase response rapidly oscillates with PA-location changes and the optimization variables are strongly coupled.
    To avoid getting stuck in poor local optima and to characterize the optimal PASS performance, the authors of \cite{xu2025pinching} proposed a globally optimal joint transmit and pinching beamforming algorithm based on the branch-and-bound (BnB) method and McCormick relaxation. 
    Although the BnB algorithm significantly accelerates the exhaustive search and mathematically guarantees obtaining the globally optimal solution, it still has high computational complexity, which is only affordable for small-scale systems.
    \item \textbf{Suboptimal algorithms:} 
    For single-waveguide PASS, the authors of \cite{xu2025rate} maximized the received signal power by exploiting a deterministic two-stage PA-placement strategy based on analytical phase-alignment and scalable path-loss expressions. 
    To balance the throughput-fairness trade-off, the authors of \cite{tegos2025minrate} investigated the uplink PASS minimum rate maximization problem and separately updated the transmit power and PA positions. 
    Successive convex approximation (SCA)-based convexification was adopted to iteratively tighten the non-convex rate constraints. 
    Closed-form and high-SNR structural solutions for max–min rate maximization, power minimization, and sum-rate maximization were derived, revealing that the optimal PA positions vary fundamentally with the underlying optimization objective. 
    For NOMA-assisted single-waveguide PASS, the authors of \cite{xuyq2025qosnoma} investigated quality-of-service (QoS)-aware downlink NOMA designs, where PA locations and power allocation coefficients were jointly optimized. By leveraging SCA and block coordinate descent (BCD), they obtained an iterative algorithm that exploited the structure of large-scale path loss and cascaded in-waveguide/free-space phase shifts, achieving near-exhaustive-search performance with substantially reduced complexity. 

    For downlink multi-waveguide PASS, the authors of \cite{wang2025modeling} formulated a joint transmit and pinching beamforming optimization problem to minimize the transmit power subject to QoS constraints, and developed penalty-based alternating optimization (AO) algorithms for both continuous and discrete activation. 
    To maximize the system sum rate, the authors of \cite{xu2025jointpass} developed a majorization–minimization (MM) and penalty dual decomposition (PDD) based algorithm for joint transmit and pinching beamforming optimization.
    Furthermore, the authors of \cite{zhang2025uplinkmiso} investigated uplink sum-rate maximization for a PASS-assisted multi-user MISO system with minimum-mean-square-error (MMSE) decoding, and developed a fractional programming-based algorithm to jointly optimize the user transmit powers and the continuous PA locations.
    The authors of \cite{zhouzw2025sumrate} considered the sum-rate maximization of NOMA-assisted PASS and derived closed-form power allocation via KKT conditions, followed by a bisection-based search for PA positions. 
    Taking into account in-waveguide attenuation and power constraints, the authors of \cite{hu2025sum} further investigated multi-waveguide NOMA-assisted PASS and applied AO combined with SCA to jointly optimize digital precoding and PA positions. 
    In \cite{zhouzh2025seee}, the authors investigated the spectral–energy efficiency trade-off and proposed iterative closed-form refinement and AO-based joint beamforming designs.

    Since the motions of PAs require extra costs, the authors of \cite{zhang2025twotimescale} developed a two-timescale joint transmit and pinching beamforming scheme, in which fast-varying transmit beamformers were optimized via a KKT-guided dual-learning approach, while slowly varying pinching beamformers were updated using stochastic SCA.  
    By explicitly modeling the motion-induced power consumption of PAs, the authors of \cite{xu2025jointradiation} investigated the joint optimization of radiation power control, PA positions, and transmit beamforming. An alternating direction method of multipliers (ADMM)-based framework was developed for continuous PA movement, whereas a BCD method was developed for discrete activation.

For broader PASS applications, the authors of \cite{sun2025pls} investigated physical-layer security (PLS) in PASS and designed secrecy-rate-maximizing beamformers by combining fractional programming and BCD. 
For symbiotic radio, the authors of \cite{wang2025pass_sr} devised an SCA-enabled pinching beamforming algorithm that maximized the primary communication rate while guaranteeing the secondary backscatter link performance. 
For energy-efficient uplink PASS communication in unmanned aerial vehicle (UAV)-enabled mobile edge computing (MEC) networks, the authors of \cite{ai2025uav} solved the resulting mixed-integer nonlinear program through BCD, convex approximation, and one-dimensional search, showing that structured decomposition substantially reduced computational burden.
\end{itemize}

\begin{table*}[t]
\centering
\caption{Summary of optimization methods for PASS.}\label{table:optimization_methods}
\resizebox{1\textwidth}{!}{
\begin{tabular}{|l|l|l|l|}
\hline
\textbf{Category} & \textbf{Method} & \textbf{Key Techniques} & \textbf{Literature} \\ \hline

\multirow{5}{*}{Structure-based Optimization}
& \textbf{Analytical optimization}
& Closed-form derivations \& structural analysis
& \cite{ding2025analytical}, \cite{xu2025rate} \\ \cline{2-4}

% Structure-based Optimization
& \textbf{Global optimization}
& Branch-and-bound (BnB)
& \cite{xu2025pinching} \\ \cline{2-4}

% Structure-based Optimization
& \textbf{Penalty-based methods}
& Penalty optimization, ADMM, PDD
& \cite{wang2025modeling}, \cite{zhang2025integrated} 
\cite{li2025pinching},
\cite{zhang2025twotimescale}, \cite{xu2025jointradiation}\\ \cline{2-4}

% Structure-based Optimization
& \textbf{Convex-approximation methods}
& SCA, MM, other convex surrogates
& \cite{tegos2025minrate}, \cite{xuyq2025qosnoma}, \cite{xu2025jointpass}, \cite{hu2025sum}, \cite{zhouzh2025seee}, \cite{wang2025pass_sr}, \cite{ai2025uav}\\ \cline{2-4}

% Structure-based Optimization
& \textbf{Fractional programming}
& Fractional programming, quadratic transforms
& \cite{zhang2025uplinkmiso} \\ \hline

\multirow{2}{*}{Population-based Optimization}
& \textbf{Particle swarm optimization}
& Swarm-based search, velocity-position updates
& \cite{wang2025sensing}, \cite{jiang2025covert}, \cite{gan2025noma,zeng2025noma} \\ \cline{2-4}

% Population-based Optimization
& \textbf{Hybrid methods}
& KKT-guided PSO, PSO-ZF, SCA-PSO
& \cite{Xu2026ModeCombiningPASS},\cite{gan2025noma}, \cite{zhu2025secure} \\ \hline

\multirow{3}{*}{Game-theoretic Optimization}
& \multirow{2}{*}{\textbf{Matching game}}
& One-to-one matching
& \cite{wangkd2025activation}  \\ \cline{3-4}

& %\textbf{Matching game}
& Many-to-many matching
& \cite{xu2025pinching}  
\\ \cline{2-4}

% Game-theoretic Optimization
& \textbf{Coalition game}
& Merge-and-split, Shapley-value evaluation
& \cite{wangkd2025multiwg,wangkd2025pls} \\ \hline

\end{tabular}}
\end{table*}

\subsubsection{Population-based Optimization Methods}
In contrast to structure-based optimization methods, \textit{population-based optimization methods} aim to explore the non-convex design space more aggressively through stochastic and metaheuristic search, such as particle swarm optimization (PSO) and differential evolution.

In existing works, PSO has been widely used for PASS optimization. 
For PASS-based wireless sensing using leaky coaxial cables for echo-signal reception, the authors of \cite{wang2025sensing} formulated a CRLB minimization problem and proposed a two-stage PSO algorithm to jointly optimize the transmit waveform and PA positions.
For covert communication scenarios, the authors of \cite{jiang2025covert} derived closed-form optimal positions for a single PA, and then extended it to more general configurations via a two-stage PSO procedure, where particles represent candidate PA positions and fitness was defined by the achievable covert rate. 
In NOMA-assisted PASS, the authors of \cite{gan2025noma} developed both MM-based and PSO-based algorithms. 
The structured MM-PDD algorithm served as a baseline, while the PSO-ZF scheme constructed particles that encode PA positions and power allocation, embedding a ZF-based transmit beamforming rule for fast fitness evaluation. Numerical results demonstrated that the PSO-ZF approach can escape the unfavorable local minima encountered by MM-based optimization \cite{gan2025noma}. 
Furthermore, the authors of \cite{zeng2025noma} investigated uplink NOMA-assisted single-waveguide PASS and developed PSO algorithms to jointly optimize transmit powers and PA positions. 
For secure PASS-enabled transmissions, the authors of \cite{zhu2025secure} designed secrecy-rate-maximizing strategies by combining SCA and PSO for joint beamforming optimization. 
To further reduce the search space of PSO for fast convergence, the authors of \cite{Xu2026ModeCombiningPASS} proposed a KKT-guided PSO algorithm for the joint PA configuration and transmit beamforming optimization in M-PASS, which exploited KKT-conditioned beamforming solutions to guide the particle-swarm search. 

\subsubsection{Game-theoretic Methods}
Game-theoretic methods reformulate the PASS optimization problem into matching or coalitional games, thus explicitly modeling the interactions among multiple PAs, waveguides, and users. Then, algorithmic mechanisms can be designed to obtain stable matchings or coalitions. 
\begin{itemize}
    \item \textbf{Matching game-based methods:} For a single waveguide PASS, the authors of \cite{wangkd2025activation} tackled the antenna activation in NOMA-assisted PASS and recast the optimization of the locations and the number of activated PAs as a one-sided one-to-one matching problem. 
    A low-complexity matching algorithm was derived. Simulation results confirmed that the matching-based design yielded near-optimal throughput with low complexity. For multi-waveguide PASS, the authors of \cite{xu2025pinching} reformulated the joint discrete pinching beamforming and transmit beamforming problem as a many-to-many matching game between waveguides and candidate PA positions. A welfare-driven matching algorithm was proposed, which was shown to converge to pairwise-stable solutions with a performance close to the globally optimal BnB. 
    \item \textbf{Coalition game-based methods:} In \cite{wangkd2025multiwg}, the authors jointly optimized waveguide assignment, antenna activation, SIC decoding order, and power allocation by treating waveguide assignment and antenna activation as separate coalition-formation games. Coalition formation captured the cooperation among PAs and users, while monotonic optimization and SCA were used for power-control subproblems, yielding a hybrid game-theoretic/structured-deterministic framework. For PLS-oriented designs, the authors of \cite{wangkd2025pls} further modeled the cooperation among PAs as a coalitional game with non-transferable utility, where each PA's contribution was quantified via the Shapley value or marginal contribution. A coalitional game-based antenna activation algorithm was then proposed to enhance the secrecy rate under practical discrete deployment constraints. 
\end{itemize}

Existing studies of PASS optimization generally fall into one of the above three categories. Specifically, structure-based methods are preferred when accurate models and tractable surrogates are available. Population-based methods can be employed to enhance performance for moderate-scale scenarios with higher computation costs. 
Game-theoretic approaches, such as matching and coalitional games, offer a principled way to design low-complexity algorithms with stability guarantees for discrete activation and resource allocation in practical large-scale PASS.

\subsection{A Machine Learning Perspective}
Despite the many numerical algorithms proposed for PASS, optimizing large-scale systems remains challenging due to the non-convex nature of the spherical-wave channel. This makes it difficult to find optimal or near-optimal solutions to these problems efficiently. For instance, exhaustive search can be used to find the optimal PA positions, but this method is computationally expensive when applied to many PAs. 
Thus, this method cannot meet the real-time implementation requirements of practical systems.

To address this challenge, deep learning has been introduced into PASS to learn the optimal positions and activation patterns of PAs along with corresponding power allocations. Compared to optimization-based methods, the deep-learning-based approach enables real-time implementation with low computational complexity, especially for joint optimization problems, such as optimizing PA positions and beamforming. Furthermore, since the optimization problems in PASS are usually non-convex, the deep-learning-based approach is expected to outperform optimization-based methods.

\subsubsection{Learning for Communications}
In \cite{karagiannidis2025deep}, an antenna activation scheme in a single-waveguide PASS was learned to maximize SE. A graph neural network (GNN) was designed and used for learning, motivated by its ability to capture spatial and connectivity patterns of wireless systems and generalize effectively across different configurations. Simulation results demonstrated that the proposed GNN generalizes well to large-scale antenna arrays and is robust to user localization uncertainties.

\begin{figure}[!t]
    \centering
    \includegraphics[width=0.75\linewidth]{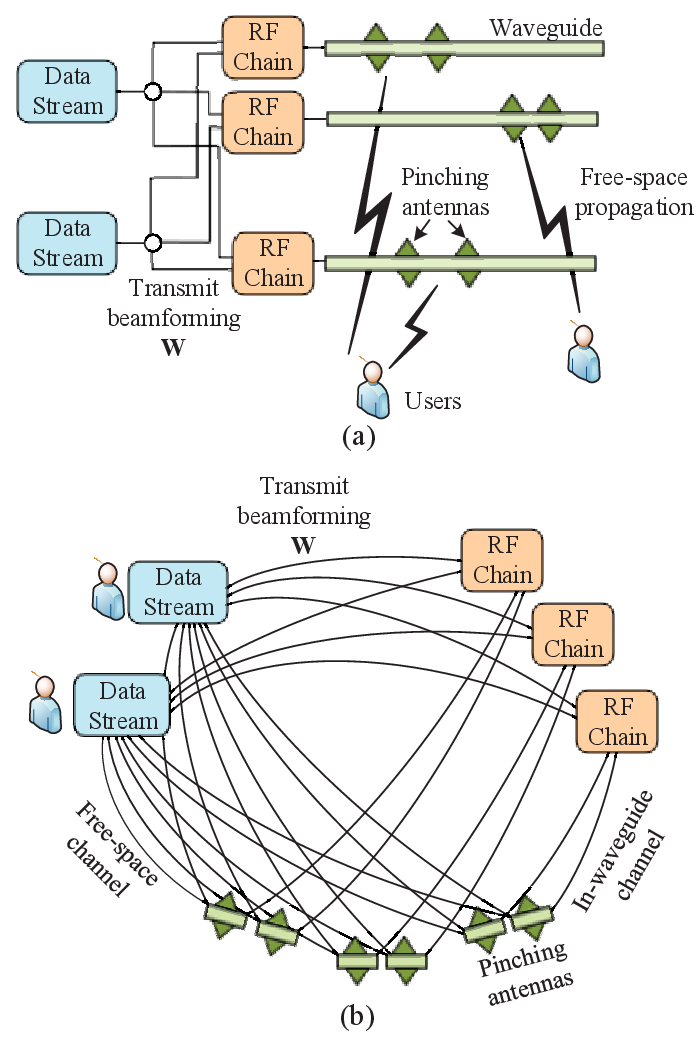}
    \caption{Illustration of the system model and the corresponding graph representation for learning-based joint PA-positioning and transmit beamforming design. 
(a) A multi-waveguide PASS downlink system, where data streams are mapped to RF chains through the transmit beamforming matrix $\mathbf{W}$, propagated along the waveguides, radiated by the PAs, and then received by the users through free-space links. 
(b) The corresponding heterogeneous graph, where different types of nodes represent data streams, RF chains and PAs, while different types of edges characterize transmit beamforming links, in-waveguide channels, and free-space channels.}
    \label{fig:graph}
\end{figure}

For jointly learning PA positions and other transmission strategies, a GNN architecture was also proposed in \cite{GJ-WCL} to jointly learn antenna positions and transmit beamforming matrices for maximizing SE. This work is among the earliest studies on learning-aided PASS. The system model and the modeled graph that the GNN learns on is shown in Fig. \ref{fig:graph}. The GNN employs a staged architecture of first learning PA positions and then learning the transmit beamforming matrix when the positions are given. For the first time, it was revealed that the joint policy was not affected by permuting the users and waveguides. The permutation property was leveraged when designing the GNN to enable scalability, i.e., low training complexity even for large-scale problems. Simulation results demonstrated that the proposed GNN can learn a near-optimal joint policy with low computational complexity. 
This work was then extended to learning tri-hybrid beamforming in \cite{guo2026graph}.
Besides, GNN is also exploited in understanding environmental information, such as a site-specific learning framework in \cite {jia2026spefic}.

The joint learning of beamforming and antenna positions was also studied in \cite{kang2025campass}, aiming to maximize the SE in a multi-waveguide IoT system with a single PA on each waveguide. Similar to \cite{GJ-WCL}, a multi-block-based DNN architecture was proposed for learning, where the blocks are respectively used for feature extraction, antenna positioning, and precoding design. It was demonstrated via simulations that the proposed architecture can achieve near-optimal SE.

A gradient-input unfolding network was proposed in \cite{zhou2025gradient} to jointly learn antenna positions and transmit beamforming matrices for maximizing the weighted sum rate (WSR) of multiple users. A multi-waveguide scenario with a single antenna deployed on each waveguide was again considered. The proposed DNN architecture can be regarded as a model-based deep unfolding network, where the mathematical expressions of the gradients with respect to the objective function and constraints are introduced into the DNN to simplify the mappings to be learned. 
Simulation results showed that the proposed learning-based method can achieve rapid convergence and higher WSR than an optimization-based method.

As for the multi-waveguide multi-PA scenario, the joint learning of transmit beamforming and PA positions was studied in \cite{xu2025jointpass}, which also aims to maximize SE. A Transformer-based architecture was proposed, where the KKT solution structure of transmit beamforming was leveraged, such that the DNN only needs to learn a few dual variables instead of the whole transmit beamforming matrix. Simulation results revealed that the proposed KDL-Transformer can achieve much higher SE than an optimization-based method with millisecond-level response, enabling real-time implementation.

\subsubsection{Learning for ISAC}
In addition to learning to optimize communication performance, recent works have also considered learning-based PASS designs that jointly support sensing and communication.
The joint learning of PA positions and beamforming was investigated in \cite{GJ-GC25}, aiming to maximize the sensing performance of a target while satisfying the minimum data requirements of multiple users. This work was then extended to the multi-target multi-user scenario in \cite{GJ-JSTSP,GJ-GC25}, where a GNN was proposed to leverage both the permutation property and a derived optimal solution structure for the transmit beamforming. 
Simulation results confirmed that the learning-based method can achieve performance close to or better than that of an optimization-based method \cite{gj2026icc}, with much lower computational complexity. 

Recent papers on learning-based methods for PASS are summarized in Table \ref{tab:learning}.

\begin{table*}[t]
\centering
\caption{Summary of references on learning-based methods for PASS.}
\label{tab:learning}

\footnotesize
\setlength{\tabcolsep}{3pt}
\renewcommand{\arraystretch}{1.1}

\resizebox{0.98\textwidth}{!}{%
\begin{tabular}{|c|
                >{\raggedright\arraybackslash}m{4.4cm}|
                >{\raggedright\arraybackslash}m{3.3cm}|
                >{\raggedright\arraybackslash}m{2.6cm}|
                >{\raggedright\arraybackslash}m{3.2cm}|
                c|c|}
\hline
\textbf{Ref.} &
\textbf{Motivation of Learning} &
\textbf{Policy} &
\textbf{Objective} &
\textbf{DNN architecture} &
\shortstack{Permutation\\property used?} &
\shortstack{Model-\\based?} \\
\hline

\cite{karagiannidis2025deep} &
\shortstack[l]{Universal approximation ability,\\low implementation complexity} &
PA activation &
Maximize SE &
GNN &
\ding{51} &
\ding{55} \\
\hline

\cite{GJ-WCL} &
\shortstack[l]{Universal approximation ability,\\low implementation complexity} &
\shortstack[l]{PA positions and\\transmit beamforming} &
Maximize SE &
GNN &
\ding{51} &
\ding{51} \\
\hline

\cite{zhou2025gradient} &
\shortstack[l]{Better performance than optimization\\-based method for non-convex problems} &
\shortstack[l]{PA positions and\\transmit beamforming} &
Maximize weighted sum-rate &
\shortstack[l]{Gradient-input unfolding\\network} &
\ding{55} &
\ding{51} \\
\hline

\cite{kang2025campass} &
\shortstack[l]{Effective at solving problems\\that are not mathematically tractable} &
\shortstack[l]{PA positions and\\transmit precoding} &
Maximize SE &
\shortstack[l]{Multi-block based\\architecture} &
\ding{55} &
\ding{55} \\
\hline

\cite{xu2025jointpass} &
\shortstack[l]{Low implementation\\complexity} &
\shortstack[l]{PA positions and\\transmit beamforming} &
Maximize SE &
KDL-Transformer &
\ding{55} &
\ding{51} \\
\hline

\cite{GJ-GC25,GJ-JSTSP} &
\shortstack[l]{Low implementation\\complexity} &
\shortstack[l]{PA positions and\\transmit beamforming} &
Maximize sensing rate &
GNN &
\ding{51} &
\ding{51} \\
\hline
\end{tabular}}
\end{table*}

In summary, learning-based methods are superior to optimization-based methods in the following aspects:
\begin{itemize}
\item The DNNs have the capacity of approximating any Lipschitz-continuous functions, thus have the potential of learning better decisions than human-designed numerical algorithms from data, even if the loss functions for training the DNNs are non-convex \cite{Eisen}. This has been demonstrated by the results in \cite{kang2025campass,xu2025jointpass,GJ-JSTSP}.
\item Once trained, the DNN has much lower computational complexity in the inference phase than the optimization-based methods. This is because a DNN with multiple layers can be regarded as an algorithm with multiple iterations \cite{GJ-attention,GJ-RGNN}. By learning from data, the DNN has the potential to obtain a better descent direction or step size, such that the DNN needs fewer iterations (i.e., layers) for convergence.
\end{itemize}

Various learning mechanisms and DNN architectures can be designed and used in PASS. Specifically,
\begin{itemize}
    \item \emph{DNN architectures}: To enable generalizability and scalability to large problem scales, GNNs can be designed by leveraging the permutation properties of the policies to be learned \cite{GJ-WCL, GJ-GNN}. To reduce the training overhead for learning complex policies, mathematical models can be introduced into DNNs to design model-based DNNs. 
    \item \emph{Learning mechanisms}: For learning an instantaneous decision that only depends on the current state, i.e., user positions, unsupervised learning can be leveraged to avoid generating labels. For learning from a problem that can be modeled as a Markov decision process, reinforcement learning can be applied to continuously interact with environments. To enable generalizability to environmental parameters, the meta-learning approach can be used to learn a priori knowledge from various tasks.
\end{itemize}

\subsection{Discussion and Outlook}
Despite the advantages of deep-learning-based methods, learning-based methods may not be preferable to traditional methods for every PASS problem. 
To be more specific, learning-based methods are better suited for learning continuous decisions than discrete decisions. When it comes to learning a policy from an integer programming problem, a common practice is to relax the problem to a continuous optimization problem, which may cause a loss of optimality.

The antenna activation problem in PASS can be divided into continuous activation and discrete activation \cite{liu2025pinching}. 
For the continuous case, the PAs can be positioned anywhere on the waveguide, as long as the distance between the PAs is not too small to avoid mutual coupling. 
As for the discrete case, the PAs can only be placed (or activated) at predefined positions. The PA optimization problems in these two schemes are, respectively, continuous and discrete. 
Hence, learning-based methods are more suitable for the design of continuous activation schemes. 
For the design of discrete activation schemes, since the positions of PAs exhibit a relatively strong pattern, e.g., placement at positions close to users, traditional searching algorithms, e.g., one-dimensional search or matching-based methods, already perform well and are of low complexity. Hence, a hybrid learning scheme can be adopted, where traditional search methods are used for optimizing PA positions and  learning-based methods are used to optimize other transmission strategies, e.g., transmit beamforming.

Moreover, in high-frequency channels, the learning-based methods are  constrained by the short channel coherence time. The offline training and low-complexity online inference can be performed over slowly varying variables, such as blockage-aware PA placement/activation and user association. In comparison, fast-timescale variables, e.g., baseband beamforming, need to be handled by lightweight model-based optimization.

A comparison between optimization-based and learning-based methods is provided in Table \ref{tab:comparison}.

\begin{table*}[!ht]
\centering
\renewcommand{\arraystretch}{1.1}
\caption{Comparison between optimization-based and learning-based methods for PASS}\label{tab:comparison}
\footnotesize
\setlength{\tabcolsep}{4pt}

\begin{tabular}{|c|
>{\raggedright\arraybackslash}p{3.8cm}|
>{\raggedright\arraybackslash}p{3.5cm}|
>{\raggedright\arraybackslash}p{2.8cm}|
>{\raggedright\arraybackslash}p{3cm}|}
\hline
& \textbf{Main techniques / DNN architectures} & \textbf{Advantages} & \textbf{Disadvantages} & \textbf{Recommended scenarios} \\
\hline
Optimization-based
& Structured deterministic optimization, global search, and game-theoretic formulations.
& Able to find optimal PA positions for discrete activation schemes via global search.
& High computational complexity when searching for optimal or near-optimal solutions for large-scale problems.
& Suitable for discrete activation schemes for small-scale problems. \\
\hline
Learning-based
& Algorithms such as deep learning, reinforcement learning, and meta-learning, with architectures including GNN and model-based DNN.
& Lower computational complexity and potential to provide better decisions for non-convex problems.
& Less suitable for discrete optimization problems.
& Suitable for continuous activation schemes for large-scale problems. \\
\hline
\end{tabular}
\end{table*}

\section{Conclusions} \label{sect:conclusions}
This paper provided a comprehensive survey of PASS, encompassing its physical foundations, newly emerging variants, PASS-assisted sensing, performance analysis, and mathematical tools for PASS optimization.
Specifically, the implementation of PASS was examined from the perspectives of EM theory, coupling theory, and signal modeling, which serve as its foundations.
A wide range of newly emerged PASS variants, including SWAN, C-PASS, and M-PASS, were examined with respect to their modeling, benefits, and characteristics.
To enable another important functionality, PASS-assisted sensing was also discussed, underscoring its benefits, application scenarios, and sensing-communication interplay.
Furthermore, a comprehensive survey of PASS-based communication and sensing performance analysis was presented, followed by a section shedding light on advanced mathematical tools for PASS optimization, ranging from conventional approaches to machine learning.
Finally, we identify open problems and outline future research directions within the realm of PASS in what follows:

\subsubsection{Channel estimation in PASS} 
Channel estimation constitutes a critical yet demanding task for PASS design \cite{xiao2025channel,zhou2025channel,liu2025joint}, and it is fundamentally impeded by two hardware-induced constraints. The first challenge arises from the single-RF receiver architecture on one waveguide. In this setup, the aggregation of signals from multiple distributed PAs into a single RF chain creates a ``many-to-one'' mapping, which results in compressed observation dimensions and a severely rank-deficient sensing matrix. Second, the channel characteristics are intrinsically coupled with the hardware configuration, specifically the PA positions and power radiation ratios. Consequently, frequent dynamic adjustments of these parameters would necessitate repetitive channel re-estimation, incurring prohibitive pilot overhead and latency.

Although initial studies have addressed channel estimation for PASS, a general acquisition framework remains open for channels with hardware-coupled coefficients and dimension-compressed observations. Future research may develop structure-aware estimation principles that exploit the geometric, propagation, and hardware characteristics of PASS, rather than relying on direct full-dimensional CSI acquisition. Integrating PASS-assisted sensing and localization with channel acquisition is also promising, since environmental and position information can support efficient CSI inference. Robust estimation under hardware calibration errors, PA positioning uncertainty, mutual coupling, and waveguide-related impairments is another key direction for practical deployment.

\subsubsection{Reciprocity of PAs} The reciprocity of PAs is another open problem for PASS-based wireless communications. In general, there are two possible PA options, namely reciprocal PAs and non-reciprocal PAs. For reciprocal PAs, bidirectional transmission is possible, i.e., allowing signals to not only be radiated from the waveguide into the free space but also be received from the free space into the waveguide. As a result, full-duplex transmission is facilitated by reciprocal PAs, which is useful for advanced designs, such as ISAC and MEC. Nevertheless, it should be noted that reciprocal PAs may lead to the inter-PA radiation issue during signal reception at multiple PAs, i.e., signals received by one PA can be further re-radiated by other PAs during in-waveguide propagation. This degrades the received signal strength in the baseband and further complicates CSI acquisition. For non-reciprocal PAs, a strict directionality is achieved by allowing signal radiation either solely
from the free space to the waveguide, termed uplink PAs, or from the waveguide into free space, termed as downlink PAs. By doing so, inter-PA radiation in the uplink can be resolved. However, non-reciprocal PAs can only support half-duplex operation and require PASS to employ two different PAs for downlink and uplink transmission, which increases hardware complexity. Therefore, the investigation of performance and complexity trade-offs achieved by reciprocal PAs and non-reciprocal PAs constitutes an interesting future research direction.

Although reciprocal and non-reciprocal PAs have been discussed as candidate architectures, their system-level implications are not yet fully understood. A key open issue is to establish reciprocity-aware models that jointly capture bidirectional signal access, inter-PA reradiation, and the resulting trade-off among uplink reception, full-duplex or ISAC functions, and hardware complexity. Future designs may move toward hybrid or controllable PA reciprocity, where PA directionality is matched to the communication and sensing requirements of PASS.

\subsubsection{Implementation Challenges and Practical Issues}
Regarding practical implementation, the EM performance of PASS is critically dependent on the physical interface between the PAs and the waveguide. Specifically, geometric parameters such as the contact shape, dimensions, and mounting positions play a decisive role in shaping the signal radiation characteristics, including the radiated power intensity, beam directionality, and the 3dB beamwidth (or half-power beamwidth). Consequently, tailoring the PA design to meet specific scenario requirements is of paramount importance. For instance, a conformal annular PA architecture, which wraps circumferentially around a waveguide segment, holds the potential to achieve $360^\circ$ omnidirectional coverage. Nevertheless, bridging the gap between theoretical modeling and physical deployment for such complex geometries poses substantial challenges, warranting extensive research efforts in the future.

Although several PA implementation models have been reported, a calibrated bridge between EM-level device behavior and system-level optimization remains lacking. Future research should integrate EM design, hardware characterization, and communication-theoretic optimization so that radiation behavior, coupling efficiency, and mounting conditions can be represented by tractable system parameters. Hardware-constrained PA placement, reconfiguration, and resource allocation should also be studied together with scalability, reliability, maintenance, and wideband operation to guide practical PASS deployment.

\subsubsection{Outdoor PASS Deployments and Hybrid RF/FSO Fronthaul}
Outdoor deployment is also an important future direction for PASS, since it features large-scale channel reconfigurability to mitigate path losses and LoS link creation to build robust communication links between transceivers.
In current PASS architectures, the waveguides are connected to the BS via wired links, which limits the flexibility of PASS deployment.
To address this issue, one approach is to use Wi-PASS \cite{wijewardhana2025wireless}, which replaces wired connections with wireless counterparts.
In parallel, another approach is to utilize hybrid RF/free-space optical (RF/FSO) links as a promising wireless fronthaul or backhaul solution.
In particular, FSO links can provide high-capacity directional transmission, while RF links can maintain robust connectivity under adverse weather and misalignment conditions \cite{khalighi2014survey}.
By combining RF and FSO links, outdoor PASS deployments can benefit from both robust connectivity and high-capacity directional transmission.

\section*{Acknowledgment}
Yuanwei Liu led and supervised the whole processing of this work.
Hao Jiang led the literature review, organized the technical content of the survey, prepared the original manuscript draft, and coordinated the revisions. 
Xu Gan, Xiaoxia Xu, Jia Guo, Zhaolin Wang, Chongjun Ouyang, and Xidong Mu contributed to the technical development and verification of the literature review and provided input on the discussion of PASS modeling, architectures, communication, sensing, and emerging applications. 
Zhiguo Ding, Arumugam Nallanathan, Octavia A. Dobre, George K. Karagiannidis, and Robert Schober provided technical guidance, critical feedback, and substantial revisions to the manuscript.

\bibliographystyle{IEEEtran}
\bibliography{mybib}

\end{document}